\documentclass{royal1}
\renewcommand{\thesection}{\arabic{section}}
\renewcommand{\thefigure}{\arabic{figure}}
\begin{document}
\chapter{Bloch oscillations and Wannier-Stark localization 
in semiconductor superlattices}
\vspace*{4cm}
\begin{center}
{\large
Fausto Rossi \\[.5cm]
{\it 
Istituto Nazionale per la Fisica della Materia \\
Dipartimento di Fisica, Universit\`a di Modena \\
Via G. Campi 213/A, I-41100 Modena, Italy
}
}
\end{center}
\newcommand\apt{&\!\!\!\!\!\!\!\!\!\!&}
\newcommand\prr[1]{\protect\ref{#1}\protect}
\newcommand\prc[1]{\protect\cite{#1}\protect}
\section{Introduction}\label{s:int}

Ever since the initial applications of quantum mechanics to the dynamics of
electrons in solids, the analysis of Bloch electrons moving in a 
homogeneous electric field has been of central importance.

By employing semiclassical arguments, in 1928 Bloch \prc{Bloch28} 
demonstrated that, a wave packet given by a superposition of single-band states 
peaked about some quasimomentum, $\hbar{\bf k}$, moves 
with a group velocity given by the gradient of the energy-band function 
with respect to the quasimomentum and that the rate of change of the 
quasimomentum is proportional to the applied field ${\bf F}$. 
This is often referred to as the ``acceleration theorem'':
\begin{equation}\label{eq0}
\hbar\dot{\bf k} = e{\bf F}\ .
\end{equation}
Thus, in the absence of interband tunneling and scattering processes, 
the quasimomentum of a Bloch 
electron in a homogeneous and static electric field will be uniformly 
accelerated into the next Brillouin zone in a repeated-zone scheme (or 
equivalently undergoes an Umklapp process back in to the first zone). 
The corresponding motion of the Bloch electron through the periodic 
energy-band structure, shown in Fig.~\prr{fig1}, is called ``Bloch 
oscillation''; It is characterized 
by an oscillation period $\tau_B = h/eFd$, where $d$ denotes the lattice 
periodicity in the field direction. 

There are two mechanisms impeding a fully periodic motion: interband 
tunneling and scattering processes.
Interband tunneling is an intricate problem and still at the center of a 
continuing debate.
Early calculations of the tunneling probability into other bands in which 
the electric field is represented by a time-independent scalar potential 
were made by Zener \prc{Zener34} using a Wentzel-Kramers-Brillouin 
generalization of Bloch functions, 
by Houston \prc{Houston40} using accelerated Bloch states (Houston 
states),
and subsequently by Kane \prc{Kane59} and Argyres \prc{Argyres62} 
who employed the crystal-momentum representation.
Their calculations lead to the conclusion that the tunneling 
rate per Bloch period is much less than unity for electric fields up to 
$10^6$V/cm for 
typical band parameters corresponding to elemental or compound 
semiconductors.

Despite the apparent agreement among these calculations, the 
validity of employing the crystal-momentum representation or Houston 
functions to describe electrons moving in a 
non-periodic (crystal plus external field) potential has been 
disputed. 
The starting point of the controversy was the original paper by 
Wannier \prc{Wannier60}. 
He pointed out that, 
due to the translational symmetry of the crystal potential, 
if $\phi({\bf r})$ is an 
eigenfunction of the scalar-potential Hamiltonian (corresponding to the 
perfect crystal plus the external field)
with eigenvalue 
$\epsilon$, 
then any $\phi({\bf r}+n{\bf d})$ is also an
eigenfunction with eigenvalue 
$\epsilon+n \Delta\epsilon$, where $\Delta\epsilon = e F d$ is the 
so-called Wannier-Stark splitting
(${\bf d}$ being the primitive 
lattice vector along the field direction). 
He concluded that the translational symmetry of the crystal gives rise
to a discrete energy spectrum, the so-called Wannier-Stark ladder. The 
states corresponding to these equidistantly spaced levels are localized 
states, as schematically shown in Fig.~\prr{fig2} for the case of a 
semiconductor superlattice.
The degree of this Wannier-Stark 
localization depends on the strength of the applied field.

The existence of such energy quantization was disputed 
by Zak \prc{Zak68a}, who pointed out that 
for the case of an infinite crystal the scalar potential 
$-{\bf F}\cdot{\bf r}$ is not bounded, which implies a continuous energy 
spectrum.
Thus, the main point of the controversy was related to the 
existence (or absence) of Wannier-Stark ladders. 
More precisely, the point 
was to decide if interband tunneling (neglected in the original calculation
by Wannier \prc{Wannier60})
is so strong to destroy the 
Wannier-Stark energy quantization (and the corresponding Bloch oscillations) 
or not.
We will give a brief historical account of this 
long-standing controversy in the following section.

It is only during the last decade that this controversy came to an end.
From a theoretical point of view, most of the formal problems 
related to the non-periodic nature of the scalar potential (superimposed to 
the periodic crystal potential) were finally removed by using 
a vector-potential representation of the applied 
field \prc{Kittel63,Krieger86}.
Within such vector-potential picture, 
upper boundaries for the interband 
tunneling probability have been established at a rigorous level, which show
that an electron may execute a number of Bloch oscillations before 
tunneling out of the band \prc{Krieger86,Nenciu91,DiCarlo94}, in 
qualitatively good agreement with the earlier predictions of Zener and Kane 
\prc{Zener34,Kane59}.

The second mechanism impeding a fully periodic motion is scattering by 
phonons, impurities, etc. (see Fig.~\prr{fig1}). 
This results in lifetimes shorter than the Bloch 
period $\tau_B$ for all reasonable values of the electric 
field, so that Bloch oscillations should not be observable in conventional 
solids. 

In superlattices, however, the situation is much more favourable because of 
the smaller Bloch period $\tau_B$ resulting from the small width of the 
mini-Brillouin zone in the field direction \prc{Bastard89}.

Indeed, the existence of Wannier-Stark ladders as well as 
Bloch oscillations in 
superlattices has been confirmed by a number of recent 
experiments \prc{Shah96}.
The photoluminescence and photocurrent measurements of the biased 
GaAs/GaAlAs superlattices performed by Mendez and coworkers 
\prc{Mendez88},
together with the electroluminescence experiments by Voisin and 
coworkers \prc{Voisin88},
provided the earliest experimental
evidence of the field-induced Wannier-Stark ladders in superlattices. 
A few years later, Feldmann and coworkers \prc{Feldmann92} were able to 
measure Bloch oscillations in the time domain through a 
four-wave-mixing experiment originally suggested by 
von Plessen and Thomas \prc{vonPlessen92}.
A detailed analysis of the Bloch oscillations in the four-wave-mixing signal 
(which reflects the interband dynamics) has been also performed by Leo and 
coworkers \prc{Leo92,Leisching94}.

In addition to the above interband-polarization analysis, Bloch 
oscillations have been also detected by monitoring the intraband 
polarization which, in turn, is reflected by anisotropic changes in the 
refractive index \prc{Shah96}. 
Measurements based on transmittive electrooptic sampling (TEOS) have been 
performed by Dekorsy and coworkers \prc{Dekorsy94,Dekorsy95}.  
Finally Bloch oscillations were recently measured through a direct 
detection of the TeraHertz (THz) radiation in 
semiconductor superlattices \prc{Waschke93,Roskos94}.

The aim of this chapter is to present a general approach to the study of 
the ultrafast carrier dynamics in semiconductor superlattices. 
Our theoretical description, based on the 
density matrix formalism discussed in chapter 6, is presented in 
Sect.~\prr{s:ta}.
It allows to derive a set of kinetic 
equations which accounts for interband tunneling as well as scattering 
processes and it is valid in any quantum-mechanical representation. 

In Sect.~\prr{s:qmp} the freedom of choice of the basis 
states in our kinetic formulation will be used to introduce the two typical 
pictures commonly used for the 
description of semiconductor superlattices, 
namely, the Bloch-oscillation and Wannier-Stark pictures.
In particular, we will see that they correspond to the two equivalent 
vector- and scalar-potential representations of the applied field.
This will implicitly state the total equivalence of the Bloch-oscillation and 
Wannier-Stark representations,
which in turn shows that the 
so-called ``semiclassical Bloch picture'' is on the contrary a rigorous 
quantum-mechanical result.

Finally, in Sect.~\prr{s:sse} we will review 
and discuss some simulated experiments. 
\section{Hystorical background}\label{s:hb}

In this section we give a brief historical account concerning the 
controversy on the existence of Wannier-Stark ladders mentioned above.
Some of the main criticisms to the pioneering works on Bloch oscillations 
may be summarized as follows.
\begin{itemize}
\item[(i) ]
The eigenvalues of the time-independent Schr\"odinger 
equation are not quantized but they form a continuous spectrum.
\item[(ii) ]
Since the Hamiltonian within the scalar-potential representation is not 
periodic, 
it is not clear whether one can employ the periodic Bloch states or 
Houston functions:
a superposition of Bloch functions will automatically yield a periodic 
function while the solution of the 
time-dependent Schr\"odinger equation is, in general, not periodic.
\item[(iii) ]
The position operator (entering the scalar potential) is hill-defined with\-in 
the crystal-momentum representation.
\end{itemize}
\par\noindent
In a series of papers, Wannier \prc{Wannier60,Wannier62a} and 
coworkers \prc{Wannier62b,Wannier68} have argued that in the 
presence of a homogeneous electric field, one can modify the Bloch 
states in such a way that there is no interband coupling and an electron
in a crystal will move within one band with its ${\bf k}$ changing in time 
according to the acceleration theorem in Eq.~(\prr{eq0}).
Furthermore, if ${\bf k}(t = 0)$ is in the direction of a 
reciprocal-lattice vector, the periodic motion in ${\bf k}$-space gives 
rise to an energy quantization with $\Delta\epsilon = e F d$ ($d = 
{2\pi\over G}$ being the lattice constant along the field direction), 
the so-called Wannier-Stark ladders.
``The basis for this idea is that energy bands arise from the translational
symmetry of the crystalline field and this symmetry is not removed 
physically by the presence of the applied field''\prc{Wannier62a}, 
i.e. the field is still periodic with the lattice period.

These arguments have been refuted by Zak \prc{Zak68a}, who shows that, 
although it immediately follows from the one-dimensional time-independent 
Schr\"odinger equation for an infinite crystal with lattice 
constant $d$, that if $\epsilon$ is
an eigenvalue, $\epsilon+n\Delta\epsilon$ is also an eigenvalue, 
the spectrum of $\epsilon$ is 
continuous with $-\infty < \epsilon < +\infty$, so the Wannier-Stark ladders 
do not exist.

Wannier \prc{Wannier69} has argued that Zak's criticisms of his proof 
are not valid, but concedes that the Stark ladders may be metastable 
resonant states limited 
by interband tunneling, as for the case of the 
hydrogen atom in the presence of a static field.
However, Wannier's arguments were immediately rejected by Zak \prc{Zak69}, 
who claims that Wannier's original equation was incorrect. 

A few years later, 
Rabinovitch and Zak \prc{Rabinovitch72} have extended Zak's \prc{Zak68a} 
earlier arguments to the question 
of Bloch oscillations.
They argue that since neglecting the interband coupling terms in the 
time-independent Schr\"o\-din\-ger equation leads incorrectly to 
energy quantization, then the interband terms cannot be neglected in the 
lowest approximation because they are the same order as the terms retained.
By applying the same reasoning to the time-dependent equation, they conclude 
(without offering a proof) that neglecting the interband terms as a first 
approximation, as done by Houston \prc{Houston40}, is incorrect for times 
equal to or longer than the Bloch-oscillation period $\tau_B$. 
From their conclusions it follows that the typical diagrams which are 
commonly used to portray trajectories of ${\bf k}(t)$ superimposed upon the 
energy-band structure (see Fig.~\prr{fig1}) are incorrect and 
misleading.

Nevertheless, shortly before these latter arguments appeared, experimental 
results were obtained by Koss and Lambert \prc{Koss72}, which were 
interpreted as supporting the existence of Wannier-Stark levels.
They found that the observed low-temperature optical absorption of GaAs 
in a strong electric field ($F = 10^5$V/cm) closely followed the 
theoretical predictions of Callaway \prc{Callaway63}, 
which, in turn, were based on employing Kane's 
wavefunctions and Wannier-Stark quantized energy levels.
\section{Theoretical analysis}\label{s:ta}

In this section we will try to review and discuss in a systematic way 
the basic ideas used in the theoretical analysis of semiconductor 
superlattices. As already pointed out in Sect.~\prr{s:int}, the phenomena 
under investigation, i.e. Bloch oscillations and Wannier-Stark 
localization, are peculiar of any lattice structure. Therefore, even if most 
of the results discussed in this chapter refer to semiconductor 
superlattices, the general formulation presented in this section applies to 
any crystalline structure. 
\subsection{Physical system}\label{ss:ps}

In order to study the optical and transport properties of semiconductor 
superlattices, let us consider a gas of carriers in a 
crystal under the action of an applied electromagnetic field. The carriers 
will experience their mutual interaction as well as the interaction with the 
phonon modes of the crystal.
Such physical system can be described by the following Hamiltonian:
\begin{equation}\label{eq1}
{\bf H} = {\bf H}_c + {\bf H}_p + {\bf H}_{cc} + {\bf H}_{cp} 
+ {\bf H}_{pp}\ .
\end{equation}
The first term describes the noninteracting-carrier system 
in the presence of the external electromagnetic field while the 
second one refers to the free-phonon system. The last three terms describe 
many-body contributions: they refer, respectively, to carrier-carrier, 
carrier-phonon, and phonon-phonon interactions.

In order to discuss their explicit form, let us introduce the usual 
second-quantization field operators
${\bf \Psi}^\dagger({\bf r})$ and ${\bf \Psi}({\bf r})$. They describe, 
respectively, the creation and the annihilation of a carrier in ${\bf r}$.
In terms of the above field operators the carrier Hamiltonian ${\bf H}_c$ 
can be written as
\begin{equation}\label{eq2}
{\bf H}_c = \int d{\bf r} {\bf \Psi}^\dagger({\bf r})
\left[
{\left(-i\hbar\nabla_{\bf r} -{e\over c} {\bf A}({\bf r},t)\right)^2
\over 2 m_\circ} + e\varphi({\bf r},t) + V^l({\bf r})
\right] 
{\bf \Psi}({\bf r})\ . 
\end{equation}
Here, $V^l({\bf r})$ denotes the periodic potential due to the perfect crystal 
while ${\bf A}({\bf r},t)$ and $\varphi({\bf r},t)$ denote, respectively, 
the vector and scalar potentials corresponding to the external 
electromagnetic field. Since we are interested in the electrooptical 
properties as well as in the ultrafast dynamics of photoexcited carriers, 
the electromagnetic field acting on the crystal ---and the corresponding 
electromagnetic potentials--- will be the sum 
of two different contributions: the high-frequency laser field responsible 
for the ultrafast optical excitation and the additional electromagnetic 
field acting on the photoexcited carriers on a longer time-scale.
More specifically, by denoting with the labels $1$ and $2$ these two 
contributions, we can write
\begin{equation}\label{eq3}
{\bf A}({\bf r},t) = {\bf A}_1({\bf r},t) + {\bf A}_2({\bf r},t)\ , \qquad
\varphi({\bf r},t) = \varphi_1({\bf r},t) + \varphi_2({\bf r},t)
\end{equation}
and recalling that
\begin{equation}\label{eq4}
{\bf E}({\bf r},t) = -{1\over c} {\partial\over\partial t} {\bf A}({\bf r},t)
-\nabla_{\bf r} \varphi({\bf r},t)\ , \qquad 
{\bf B}({\bf r},t) = \nabla_{\bf r}\times{\bf A}({\bf r},t)
\end{equation}
we have
\begin{equation}\label{eq5}
{\bf E}({\bf r},t) = {\bf E}_1({\bf r},t) + {\bf E}_2({\bf r},t)\ , \qquad
{\bf B}({\bf r},t) = {\bf B}_1({\bf r},t) + {\bf B}_2({\bf r},t)\ .
\end{equation}
Equation (\prr{eq4}), which gives the electromagnetic fields in terms of 
the corresponding vector and scalar potentials, 
reflects the well known gauge freedom: there is an infinite number of
possible combinations of ${\bf A}$ and $\varphi$ which give rise to the same 
electromagnetic field $\{{\bf E}, {\bf B}\}$.
We will use such freedom of choice for the laser field (term $1$): we 
assume a homogeneous (space-independent) laser field ${\bf E}_1(t)$ fully 
described by the scalar potential
\begin{equation}\label{eq6}
\varphi_1({\bf r},t) = -{\bf E}_1(t) \cdot {\bf r}\ .
\end{equation}
This assumption, which corresponds to the well known dipole approximation, is 
well justified as long as the space-scale of interest is small compared to 
the light wavelength.
The explicit form of the laser field considered in this chapter is 
\begin{equation}\label{eq7}
E_1(t) = E^+(t) + E^-(t) =
E^{ }_\circ(t) e^{i\omega_L t} +
E^*_\circ(t) e^{-i\omega_L t}\ ,
\end{equation}
where $E_\circ(t)$ is the amplitude of the light field and
$\omega_L$ denotes its central frequency.

With this particular choice of the electromagnetic potentials describing 
the laser field, the Hamiltonian in Eq.~(\prr{eq2}) can be rewritten as
\begin{equation}\label{eq8}
{\bf H}^{ }_c = {\bf H}^\circ_c + {\bf H}^{ }_{cl}\ ,
\end{equation}
where 
\begin{equation}\label{eq9}
{\bf H}^\circ_c = \int d{\bf r} {\bf \Psi}^\dagger({\bf r})
\left[
{\left(-i\hbar\nabla_{\bf r} -{e\over c} {\bf A}_2({\bf r},t)\right)^2
\over 2 m_\circ} + e\varphi_2({\bf r},t) + V^l({\bf r})
\right] 
{\bf \Psi}({\bf r})
\end{equation}
describes the carrier system in the crystal under the action of the 
electromagnetic field $2$ only, while
\begin{equation}\label{eq10}
{\bf H}^{ }_{cl} = e \int d{\bf r} {\bf \Psi}^\dagger({\bf r})
\varphi_1({\bf r},t) 
{\bf \Psi}({\bf r})
\end{equation}
describes the carrier-light (cl) interaction due to the laser photoexcitation.

In analogy with the carrier system, by denoting with 
$b^\dagger_{{\bf q},\lambda}$ and $b^{ }_{{\bf q},\lambda}$ the creation 
and annihilation operators for a phonon of mode $\lambda$ and wavevector 
${\bf q}$, the free-phonon Hamiltonian takes the form
\begin{equation}\label{eq11}
{\bf H}_p = \sum_{{\bf q}\lambda} 
\hbar\omega_{{\bf q}\lambda}
b^\dagger_{{\bf q}\lambda} b^{ }_{{\bf q}\lambda}\ ,
\end{equation}
where $\omega_{{\bf q}\lambda}$ is the dispersion relation for the phonon 
mode $\lambda$.

Let us now discuss the explicit form of the many-body contributions.
The carrier-carrier interaction is described by the two-body 
Hamiltonian
\begin{equation}\label{eq12}
{\bf H}_{cc} = {1\over 2} \int d{\bf r} \int d{\bf r}' 
{\bf \Psi}^\dagger({\bf r}) {\bf \Psi}^\dagger({\bf r}') 
V_{cc}({\bf r}-{\bf r}') 
{\bf \Psi}({\bf r}') {\bf \Psi}({\bf r}) \ ,
\end{equation}
where $V_{cc}$ denotes the Coulomb potential.

Let us now introduce the carrier-phonon interaction Hamiltonian
\begin{equation}\label{eq13}
{\bf H}_{cp} = \int d{\bf r} 
{\bf \Psi}^\dagger({\bf r}) V_{cp}({\bf r}) {\bf \Psi}({\bf r})\ ,
\end{equation}
where
\begin{equation}\label{eq14}
V_{cp} = \sum_{{\bf q}\lambda} 
\left[
\tilde{g}^{ }_{{\bf q}\lambda} b^{ }_{{\bf q}\lambda} 
e^{i{\bf q}\cdot{\bf r}}
+ 
\tilde{g}^*_{{\bf q}\lambda} b^\dagger_{{\bf q}\lambda} 
e^{-i{\bf q}\cdot{\bf r}}
\right]
\end{equation}
is the electrostatic phonon potential induced by the lattice vibrations.
Here, the explicit form of the coupling function 
$\tilde{g}_{{\bf q}\lambda}$ 
depends on the particular phonon mode $\lambda$ (acoustical, optical, etc.)
as well as on the coupling mechanism considered (deformation potential, 
polar coupling, etc.).

Let us finally discuss the phonon-phonon contribution ${\bf H}_{pp}$. The 
free-phonon Hamiltonian ${\bf H}_p$ introduced in Eq.~\prr{eq11}, 
which describes a system of noninteracting phonons, by definition 
accounts only for the harmonic part of the lattice potential. 
However, non-harmonic contributions of the 
interatomic potential can play an important role in determining the lattice
dynamics in highly excited systems \prc{Kash89}, since they are 
responsible for the decay of optical phonons into phonons of lower frequency.
In our second-quantization picture, these non-harmonic contributions 
can be described in terms of a phonon-phonon interaction which induces, 
in general, transitions between free-phonon states.
Here, we will not discuss the explicit form of the phonon-phonon 
Hamiltonian ${\bf H}_{pp}$ responsible for such a decay. 
We will simply assume that such phonon-phonon interaction is so efficient to 
maintain the phonon system in thermal equilibrium. This corresponds to 
neglect hot-phonon effects \prc{Poetz83}.

It is well known that the coordinate representation used so far is not the 
most appropriate one in describing the electron dynamics within a periodic 
crystal. On the contrary, it is in general more convenient to employ the 
representation given by the eigenstates of the noninteracting-carrier 
Hamiltonian ---or a part of it--- since it automatically accounts for 
some of the symmetries of the system.
For the moment we will simply consider an orthonormal basis set 
$\{\phi_n({\bf r})\}$ without specifying which part of the 
Hamiltonian is diagonal in such representation. This will allow us to 
write down equations valid in any quantum-mechanical representation.
Since the noninteracting-carrier Hamiltonian is, in general, a function of
time, also the basis functions $\phi_n$ may be time-dependent.
Here, the label $n$ denotes, in general, a set of discrete and/or continuous 
quantum numbers.
In the absence of electromagnetic field, the above wavefunctions will 
correspond to the well known Bloch states of the crystal and the index $n$ 
will reduce to the wavevector ${\bf k}$ plus the band index $\nu$. 
In the presence of a homogeneous magnetic
field the eigenfunctions $\phi_n$ may instead correspond to Landau states. 
Finally, for the case of a constant and homogeneous electric field, 
there exist two equivalent representations: the accelerated Bloch states 
and the Wannier-Stark picture. Such equivalence results to be of crucial 
importance in understanding the relationship between Bloch oscillations and
Wannier-Stark localization and, for this reason, it will be discussed in
more detail in Sect.~\prr{s:qmp}.

Let us now reconsider the system Hamiltonian introduced so far in terms of 
such $\phi_n$ representation. As a starting point, we may expand the 
second-quantization field operators in terms of the new wavefunctions:
\begin{equation}\label{eq15}
{\bf \Psi}({\bf r}) = \sum_n \phi^{ }_n({\bf r}) a^{ }_n\ , \qquad 
{\bf \Psi}^\dagger({\bf r}) = \sum_n \phi^*_n({\bf r}) a^\dagger_n\ .
\end{equation}
The above expansion defines the new set of second-quantization operators 
$a^\dagger_n$ and $a^{ }_n$; They describe, respectively, the creation and 
annihilation of a carrier in state $n$.

For the case of a semiconductor crystal (which will be the only one considered 
in this chapter), the energy spectrum of the 
noninteracting-carrier Hamiltonian in Eq.~(\prr{eq9}) ---or a part of it---
is always characterized by two well-separated energy regions: the valence 
and the conduction band. 
Also in the presence of an applied electromagnetic field, the periodic 
lattice potential $V^l$ gives rise to a large energy gap. Therefore, we 
deal with two energetically well-separated regions, which suggests the 
introduction of the so-called electron-hole picture. 
This corresponds to a separation of the set of states $\{\phi_n\}$ 
into conduction states $\{\phi^e_i\}$ 
and valence states $\{\phi^h_j\}$.
Thus, also the creation (annihilation) operators 
$a^\dagger_n$ ($a^{ }_n$) introduced in Eq.~(\prr{eq15}) will be 
divided into creation (annihilation) electron and hole operators: 
$c^\dagger_i$ ($c^{ }_i$) and $d^\dagger_j$ ($d^{ }_j$).
In terms of the new electron-hole picture, the expansion in 
Eq.~(\prr{eq15}) is given by:
\begin{eqnarray}\label{eq16}
{\bf \Psi}({\bf r}) = \apt \sum_i \phi^e_i({\bf r}) c^{ }_i 
+ \sum_j \phi^{h *}_j({\bf r}) d^\dagger_j \nonumber \\[1ex]
{\bf \Psi}^\dagger({\bf r}) = \apt \sum_i \phi^{e *}_i({\bf r}) c^\dagger_i 
+ \sum_j \phi^h_j({\bf r}) d^{ }_j \ .
\end{eqnarray}
If we now insert the above expansion into Eq.~(\prr{eq9}), the 
noninteracting-carrier Hamiltonian takes the form
\begin{equation}\label{eq17}
{\bf H}^\circ_c = \sum_{ii'} \epsilon^e_{ii'} c^\dagger_i c^{ }_{i'} + 
\sum_{jj'} \epsilon^h_{jj'} d^\dagger_j d^{ }_{j'}
= {\bf H}^\circ_e + {\bf H}^\circ_h\ ,
\end{equation}
where 
\begin{equation}\label{eq18}
\epsilon^{e/h}_{ll'} = \pm
\int d{\bf r} \phi^{e/h *}_l({\bf r}) \left[
{\left(-i\hbar\nabla_{\bf r} -{e\over c} {\bf A}_2\right)^2
\over 2 m_\circ} + e\varphi_2 + V^l 
- \epsilon_\circ
\right] \phi^{e/h}_{l'}({\bf r})
\end{equation}
are just the matrix elements of the Hamiltonian in the 
$\phi$-representation. The  $\pm$ sign refers, respectively, to electrons 
and holes while $\epsilon_\circ$ denotes the conduction-band 
edge.
Here, we neglect any valence-to-conduction band coupling due to the 
external electromagnetic field and vice versa. 
This is well fulfilled for the systems and field-regimes  we are going to 
discuss in this chapter.
As already pointed out, the above Hamiltonian may be time-dependent. 
We will discuss this aspect in the following section, where 
we will derive our set of kinetic equations.

Let us now write in terms of our electron-hole representation the 
carrier-light interaction Hamiltonian introduced in Eq.~(\prr{eq10}):
\begin{equation}\label{eq19}
{\bf H}_{cl} = -\sum\limits_{i,j} \left[
\mu^{eh}_{ij} E^-(t) c^{\dagger}_i d^{\dagger}_j
+\mu^{eh *}_{ij} E^+(t) d^{ }_j c^{ }_i
\right]\ .
\end{equation}
The above expression has been obtained within the well known rotating-wave 
approximation by neglecting intraband transitions, absent for the case of 
optical excitations.
Here, $\mu^{eh}_{ij}$ denotes the optical 
dipole matrix element between states $\phi^e_i$ and $\phi^h_j$.

Similarly, the carrier-carrier Hamiltonian (\prr{eq12}) can be rewritten as:
\begin{eqnarray}\label{eq20}
{\bf H}_{cc} = \apt \frac{1}{2}\sum\limits_{i_1i_2i_3i_4} V^{cc}_{i_1i_2i_3i_4}
c^{\dagger}_{i_1}c^{\dagger}_{i_2}c_{i_3}c_{i_4}\nonumber\\
\apt + \frac{1}{2}\sum\limits_{j_1j_2j_3j_4} V^{cc}_{j_1j_2j_3j_4}
d^{\dagger}_{j_1}d^{\dagger}_{j_2}d_{j_3}d_{j_4}\nonumber\\
\apt - \sum\limits_{i_1i_2j_1j_2} V^{cc}_{i_1j_1j_2i_2}
    c^{\dagger}_{i_1}d^{\dagger}_{j_1}d_{j_2}c_{i_2},
\end{eqnarray}
where
\begin{equation}\label{eq21}
V^{cc}_{l_1l_2l_3l_4} = \int d{\bf r} \int d{\bf r}'
\phi^*_{l_1}({\bf r})\phi^*_{l_2}({\bf r}')
V^{cc}({\bf r}-{\bf r}') 
\phi_{l_3}({\bf r}')\phi_{l_4}({\bf r})
\end{equation}
are the Coulomb matrix elements within our $\phi$-representation.
The first two terms describe the repulsive electron-electron and hole-hole 
interaction while the last one describes the attractive electron-hole 
interaction.
Here, we neglect terms that do not conserve the number of electron-hole 
pairs, i.e. impact-ionization and Auger-recombination 
processes \prc{Quade94}, 
as well as the interband exchange interaction. 
This monopole-monopole approximation is justified as long as the
exciton binding energy (which in semiconductors is less than $20$meV) 
is small compared to the energy gap
(which is more than $1$eV).

Finally, let us rewrite the carrier-phonon interaction Hamiltonian 
introduced in Eq.~(\prr{eq13}):
\begin{eqnarray}\label{eq22}
{\bf H}_{cp} = \apt \sum\limits_{ii',{\bf q}\lambda} 
\left[ g^{e}_{ii',{\bf q}\lambda} 
c_i^\dagger b^{ }_{{\bf q}\lambda} c^{ }_{i'} +
g^{e *}_{ii',{\bf q}\lambda} 
c^\dagger_{i'} b^\dagger_{{\bf q}\lambda} c^{ }_i \right] \nonumber \\
\apt - \sum\limits_{jj',{\bf q}\lambda} \left[
g^h_{jj',{\bf q}\lambda} 
d^\dagger_j b^{ }_{{\bf q}\lambda} d^{ }_{j'} +
g^{h *}_{jj',{\bf q}\lambda} 
d^\dagger_{j'} b^\dagger_{{\bf q}\lambda} d^{ }_j \right]
\end{eqnarray}
with
\begin{equation}\label{eq23}
g^{e/h}_{ll',{\bf q}\lambda} = 
\tilde{g}^{ }_{{\bf q}\lambda} \int d{\bf r} 
\phi^{e/h *}_l({\bf r})
e^{i{\bf q}\cdot{\bf r}}
\phi^{e/h}_{l'}({\bf r})\ .
\end{equation}
In Eq.~(\prr{eq22}) we can clearly recognize four different contributions 
corresponding to electron and hole phonon absorption and emission.
\subsection{Kinetic description}\label{ss:kd}

Our kinetic description of the ultrafast carrier dynamics in semiconductor 
superlattices, presented in this section, is based on the density-matrix 
formalism. 
Since this approach has been already reviewed and discussed in chapter 6, 
here we will simply recall in our notation the kinetic equations
relevant for the analysis of carrier dynamics in superlattices, generalizing 
the approach of chapter 6 to the case of a time-dependent 
quantum-mechanical representation.

The set of kinetic variables is the same considered in chapter 6. Given our 
electron-hole representation $\{\phi^e_i\}, \{\phi^h_j\}$, 
we will consider the intraband electron and hole single-particle density 
matrices
\begin{equation}\label{eq24}
 f^e_{ii'} = \left\langle c^\dagger_i c^{ }_{i'} \right\rangle\ , \qquad
f^h_{jj'} = \left\langle d^\dagger_j d^{ }_{j'} \right\rangle
\end{equation}
as well as the corresponding interband density matrix
\begin{equation}\label{eq25}
 p^{ }_{ji} = \left\langle d^{ }_j c^{ }_i \right\rangle\ .
\end{equation}
Here, the diagonal elements $f^e_{ii}$ and $f^h_{jj}$ correspond to 
the electron and hole distribution functions of the Boltzmann theory 
while the non-diagonal terms describe intraband polarizations. 
On the contrary, the interband density-matrix elements $p^{ }_{ji}$ describe 
interband (or optical) polarizations. 

In order to derive the set of kinetic equations, i.e. the equations of 
motion for the above kinetic variables, the standard procedure starts by 
deriving the equations of motion for the electron and hole 
operators introduced in Eq.~(\prr{eq16}):
\begin{equation}\label{eq26}
c^{ }_i = \int d{\bf r} \phi^{e *}_i({\bf r}) {\bf \Psi}({\bf r})\ , \qquad
d^{ }_j = \int d{\bf r} \phi^{h *}_j({\bf r}) {\bf \Psi}^\dagger({\bf r})\ . 
\end{equation}
By applying the Heisenberg equation of motion for the field operator 
${\bf \Psi}$, i.e.
\begin{equation}\label{eq27}
\frac{d}{dt} {\bf \Psi} = \frac{1}{i\hbar} 
\left[{\bf \Psi},{\bf H}\right]\ ,
\end{equation}
it is easy to obtain the following equations of motion:
\begin{eqnarray}\label{eq28}
\frac{d}{dt} c^{ }_i = \apt \frac{1}{i\hbar} \left[c^{ }_i,{\bf H}\right] + 
\frac{1}{i\hbar}\sum_{i'} Z^e_{ii'} c^{ }_{i'} \nonumber \\[1ex]
\frac{d}{dt} d^{ }_j = \apt \frac{1}{i\hbar} \left[d^{ }_j,{\bf H}\right] + 
\frac{1}{i\hbar}\sum_{j'} Z^h_{jj'} d^{ }_{j'}\
\end{eqnarray}
with
\begin{equation}\label{eq29}
Z^{e/h}_{ll'} = i\hbar \int d{\bf r} \left(\frac{d}{dt}
\phi^{e/h *}_l({\bf r})\right) \phi^{e/h}_{l'}({\bf r})\ .
\end{equation}
As for the case of Eq.~(\prr{eq17}), here we neglect again 
valence-to-conduction band coupling and vice versa. 
Compared to the more conventional Heisenberg equations of motion, the above
equations contain an extra-term, the last one. It accounts for the 
possible time dependence of our $\phi$-re\-pre\-sen\-ta\-tion which will 
induce 
transitions between different states according to the matrix elements 
$Z^{ }_{ll'}$.

By combining the above equations of motion with the definitions of the kinetic 
variables in Eqs.~(\prr{eq24}-\prr{eq25}), our set of kinetic 
equations can be schematically written as:
\begin{eqnarray}\label{eq30}
\frac{d}{dt} f^e_{i_1i_2} = \apt \frac{d}{dt} f^e_{i_1i_2}\Biggl|_{\bf H} + 
\frac{d}{dt} f^e_{i_1i_2}\Biggl|_\phi \nonumber \\[1ex]
\frac{d}{dt} f^h_{j_1j_2} = \apt \frac{d}{dt} f^h_{j_1j_2}\Biggl|_{\bf H} + 
\frac{d}{dt} f^h_{j_1j_2}\Biggl|_\phi \nonumber \\[1ex]
\frac{d}{dt} p^{ }_{j_1i_1} = \apt \frac{d}{dt} p^{ }_{j_1i_1}\Biggl|_{\bf H} + 
\frac{d}{dt} p^{ }_{j_1i_1}\Biggl|_\phi \ .
\end{eqnarray}
They exhibit the same structure of the equations of motion (\prr{eq28}) for the 
electron and hole creation and annihilation operators: a first term induced 
by the system Hamiltonian ${\bf H}$ (which does not account for the time 
variation of the basis states) and a second one induced by the 
time dependence of the 
basis functions $\phi$.

Let us start discussing this second term, whose explicit form is given by:
\begin{eqnarray}\label{eq31}
\frac{d}{dt} f^e_{i_1i_2}\Biggl|_\phi = \apt \frac{1}{i\hbar}
\sum\limits_{i_3i_4}\left[Z^e_{i_2i_4}\delta_{i_1i_3} -
Z^e_{i_3i_1}\delta_{i_2i_4}\right] f^e_{i_3i_4} 
\nonumber \\[1ex]
\frac{d}{dt} f^h_{j_1j_2}\Biggl|_\phi = \apt \frac{1}{i\hbar}
\sum\limits_{j_3j_4}\left[Z^h_{j_2j_4}\delta_{j_1j_3} -
Z^h_{j_3j_1}\delta_{j_2j_4} \right] f^h_{j_3j_4} \nonumber \\[1ex]
\frac{d}{dt} p^{ }_{j_1i_1}\Biggl|_\phi = \apt \frac{1}{i\hbar} 
\sum\limits_{i_2j_2}\left[Z^h_{j_1j_2}\delta_{i_1i_2} +
Z^e_{i_1i_2}\delta_{j_1j_2} \right] p^{ }_{j_2i_2} \ .
\end{eqnarray}
Such terms were not considered in chapter 6, where a time-independent 
representation has been used.
As we will see in Sect.~\prr{s:qmp}, they will 
play a central role for the description of Bloch oscillations within the 
vector-potential representation.

Let us now come to the first term. This, in turn, is the sum of different 
contributions, corresponding to the various parts of the Hamiltonian.
In particular, the total Hamiltonian can be regarded as the sum of two 
terms, a single-particle contribution plus a many-body one:
\begin{equation}\label{eq32}
{\bf H} = {\bf H}_{sp} + {\bf H}_{mb} = 
\left({\bf H}^\circ_c + {\bf H}_{cl} + {\bf H}_p\right) + 
\left({\bf H}_{cc} + {\bf H}_{cp}
+ {\bf H}_{pp}\right)\ .
\end{equation}
The explicit form of the time evolution due to the single-particle 
Hamiltonian ${\bf H}_{sp}$ (non-interacting carriers plus carrier-light 
interaction plus free phonons) is given by:
\begin{eqnarray}\label{eq33}
\frac{d}{dt} f^e_{i_1i_2}\Biggl|_{sp} = \apt \frac{1}{i\hbar} 
\bigg\{\sum\limits_{i_3i_4}\left[\epsilon^e_{i_2i_4}\delta_{i_1i_3} -
\epsilon^e_{i_3i_1}\delta_{i_2i_4}\right] f^e_{i_3i_4} \nonumber \\
\apt + \sum\limits_{j_1}\left[U_{i_2j_1} p^*_{j_1i_1}
-U^*_{i_1j_1} p^{ }_{j_1i_2}\right] \bigg\} \nonumber \\
\frac{d}{dt} f^h_{j_1j_2}\Biggl|_{sp} = \apt \frac{1}{i\hbar} \bigg\{
\sum\limits_{j_3j_4}\left[\epsilon^h_{j_2j_4}\delta_{j_1j_3} -
\epsilon^h_{j_3j_1}\delta_{j_2j_4} \right] f^h_{j_3j_4} \nonumber \\
\apt + \sum\limits_{i_1}\left[U_{i_1j_2} p^*_{j_1i_1}
-U^*_{i_1j_1} p^{ }_{j_2i_1} \right] \bigg\} \nonumber \\ 
\frac{d}{dt} p^{ }_{j_1i_1}\Biggl|_{sp} = \apt \frac{1}{i\hbar} \bigg\{
\sum\limits_{i_2j_2}\left[\epsilon^h_{j_1j_2}\delta_{i_1i_2} +
\epsilon^e_{i_1i_2}\delta_{j_1j_2} \right] p^{ }_{j_2i_2} \nonumber \\
\apt + \sum\limits_{i_2j_2} U^{ }_{i_2j_2}
\left[\delta_{i_1i_2}\delta_{j_1j_2} - f^e_{i_2i_1}\delta_{j_1j_2}
- f^h_{j_2j_1}\delta_{i_1i_2} \right] \bigg\}
\end{eqnarray}
with $U_{i_1j_1} = -\mu^{eh}_{i_1j_1} E^-(t)$.

This is a closed set of equations, 
which is a consequence of the single-particle nature 
of ${\bf H}_{sp}$. 
In addition, we stress that the structure of the two contributions 
entering Eq.~(\prr{eq30}) is very similar: 
one can include the contribution (\prr{eq31}) into Eq.~(\prr{eq33}) by 
replacing $\epsilon$ with $\epsilon + Z$. 

Let us finally discuss the contributions due to the many-body part of the 
Hamiltonian: carrier-carrier and carrier-phonon interactions (the 
phonon-phonon one is not explicitly considered here).
As discussed in chapter 6, for both interaction mechanisms one can 
derive a hierarchy of equations involving higher-order density matrices and,
in order to close such equations with respect to our set of kinetic 
variables, approximations are needed.
The lowest-order contributions to our equations of motion are given by 
first-order terms in the many-body Hamiltonian: Hartree-Fock level. Since 
we will neglect coherent-phonon states, the only Hartree-Fock 
contributions will come from carrier-carrier interaction. 
They simply result in a renormalization 
\begin{equation}\label{eq34}
\Delta\epsilon^{e/h}_{l_1l_2} =
- \sum\limits_{l_3l_4} V^{cc}_{l_1l_3l_2l_4} f^{e/h}_{l_3l_4}
\end{equation}
of the single-particle energy matrices $\epsilon^{e/h}$ 
as well as in a renormalization
\begin{equation}\label{eq36}\
\Delta U_{i_1j_1} = - \sum\limits_{i_2j_2} 
V^{cc}_{i_1j_1j_2i_2} p^{ }_{j_2i_2}
\end{equation}
of the external field $U$.
(The explicit form of the renormalization terms considered in this 
chapter accounts for the Fock contributions only, i.e. no Hartree terms. The
general structure of Hartree-Fock contributions, relevant for the case of a
strongly non-homogeneous system, is discussed in chapter 6.)

We stress that the Hartree-Fock approximation, which consists in 
factorizing average values of four-point operators into products of 
two density matrices, is independent from the quantum-mechanical 
picture.
This is a general property: any mean-field approximation gives the same 
result in different representations. The reason is that the mean-field 
operation commutes with any unitary transformation connecting different 
basis states.
It is then clear that the above kinetic equations are valid in
any quantum-mechanical representation.

All the contributions to the system dynamics discussed so far describe a 
fully coherent dynamics, i.e. no scattering processes.
In order to treat incoherent phenomena, e.g. energy relaxation and 
dephasing, one has to go one step further in the perturbation expansion 
taking into account also second-order contributions (in the perturbation 
Hamiltonian ${\bf H}_{mb}$).
The derivation of these higher-order contributions, discussed in chapter 6,
will not be repeated here. Again, as for the first-order contributions 
(Hartree-Fock terms), in order to obtain a closed set of equations (with 
respect to our set of kinetic variables (\prr{eq24}-\prr{eq25})) additional 
approximations are needed, namely
the mean-field and the Markov approximation.
As for the Hartree-Fock case, 
the mean-field approximation allows to write the various higher-order 
density matrices as products of single-particle ones. 
The Markov approximation allows to eliminate the additional higher-order 
kinetic variables, e.g. phonon-assisted density matrices, providing a 
closed set of equations still local in time, i.e. no memory effects 
\prc{Brunetti89,Rossi92,TranThoai93,Schilp94}.
This last approximation is not performed in the 
quantum-kinetic theory discussed in chapter 6
where, in addition to our single-particle 
variables, one considers two-particle and phonon-assisted 
density matrices \prc{Quade94,Schilp94}.

While the mean-field approximation is representation-independent, this is 
unfortunately not the case for the Markov limit.
This clearly implies that the validity of the Markov approximation is 
strictly related to the quantum-mechanical representation considered.
We will come back to this point in the following section where the two 
different pictures used for the study of the carrier dynamics in superlattices
are discussed.

The above kinetic description, based on intra- and interband density 
matrices, allows us to evaluate any single-particle quantity.
In particular, for the analysis of semiconductor superlattices
two physical quantities play a central role: the intra- and interband 
total (or macroscopic) polarizations:
\begin{equation}\label{eq01}
P^{e/h}(t) = \sum_{ll'} M^{e/h}_{ll'} f^{e/h}_{l'l}(t)\ , \qquad
P^{eh} = \sum_{ij} \mu^{eh}_{ij} p^{ }_{ji}(t)\ ,
\end{equation}
where $M^{e/h}$ and $\mu^{eh}$ denote, respectively, the intra- and interband 
dipole matrix elements in our 
$\phi$-re\-pre\-sen\-ta\-tion.
The time derivative of the intraband polarization $P^{e/h}$ 
describes the radiation field induced by the Bloch-oscillation dynamics 
(which for a superlattice structure is in the TeraHertz range) 
while the Fourier transform of the interband (or optical) polarization 
$P^{eh}$ 
provides the optical-absorption spectrum. 
\section{Two equivalent pictures}\label{s:qmp}

In this section we will apply the theoretical approach presented so far 
to the case of a semiconductor superlattice in the presence of an 
uniform (space-independent) electric field.
The non-interacting carriers within the superlattice crystal will then be
described by the Hamiltonian ${\bf H}^\circ_c$ in Eq.~(\prr{eq9}),
where now the electrodynamic potentials ${\bf A}_2$ and $\varphi_2$ 
(in the following simply denoted with 
${\bf A}$ and $\varphi$) correspond 
to a homogeneous electric field 
${\bf E}_2({\bf r},t) = {\bf F}(t)$.

As pointed out in Sect.~\prr{ss:ps}, the natural quantum-mechanical 
representation is given by the eigenstates of this Hamiltonian:
\begin{equation}\label{eq37}
\left[
{\left(-i\hbar\nabla_{\bf r} -{e\over c} {\bf A}({\bf r},t)\right)^2
\over 2 m_\circ} + e\varphi({\bf r},t) + V^l({\bf r})
\right] \phi_n({\bf r}) = \epsilon_n \phi_n({\bf r})\ .
\end{equation}
However, due to the gauge freedom discussed in Sect.~\prr{ss:ps},
there is an infinite number of
possible combinations of ${\bf A}$ and $\varphi$ 
---and therefore of possible Hamiltonians---
which describe the same 
homogeneous electric field ${\bf F}(t)$.
In particular, one can identify two independent choices:
the vector-potential gauge
\begin{equation}\label{eq38}
{\bf A}({\bf r},t) = -c \int_{t_\circ}^t {\bf F}(t') dt'\ , \qquad 
\varphi({\bf r},t) = 0
\end{equation}
and the scalar-potential gauge 
\begin{equation}\label{eq39}
{\bf A}({\bf r},t) = 0\ , \qquad 
\varphi({\bf r},t) = -{\bf F}(t) \cdot {\bf r}
\end{equation}
(previously employed for the description of the laser photoexcitation in 
Eq.~(\prr{eq6})).

As we will see, the two independent choices correspond, respectively, 
to the well known Bloch-oscillation and Wannier-Stark pictures. 
They simply reflect two equivalent quantum-mechanical representations and, 
therefore, any physical phenomenon can be described in both
pictures. 
\subsection{The Bloch-oscillation picture}\label{ss:bop}

The vector-potential approach presented in this section, originally 
proposed by Kittel \prc{Kittel63}, is discussed in Ref.~\prc{Krieger86}.

Within the vector-potential gauge (\prr{eq38}), the above eigenvalue 
equation (\prr{eq37}) reduces to:
\begin{equation}\label{eq40}
\left[
{\left(-i\hbar\nabla_{\bf r} -{e\over c} {\bf A}(t)\right)^2
\over 2 m_\circ} + V^l({\bf r})
\right] \phi_n({\bf r}) = \epsilon_n \phi_n({\bf r})\ .
\end{equation}
In this gauge the vector potential is space-independent but, even for the 
case of a static field (${\bf F}(t) = {\bf F}_\circ$), it is always 
time-dependent. Therefore, the above Hamiltonian (together with its 
eigenvalues and eigenfunctions) will be time dependent as well.
However, for any time $t$ we can consider its ``instantaneous'' 
eigenstates $\phi_n({\bf r},t)$. They can be easily evaluated by means of 
the following transformation:
\begin{equation}\label{eq41}
\phi_n({\bf r},t) = 
\phi^\circ_n({\bf r},t) e^{{i e\over\hbar c} \chi({\bf r},t)}
\end{equation}
with
\begin{equation}\label{eq42}
\chi({\bf r},t) = {\bf A}(t)\cdot{\bf r}\ .
\end{equation}
By applying this transformation to the eigenvalue problem in 
Eq.~(\prr{eq40}), we obtain \prc{Krieger86}:
\begin{equation}\label{eq43}
\left[
-{\hbar^2\nabla^2_{\bf r}\over 2 m_\circ} + V^l({\bf r})
\right] \phi^\circ_n({\bf r}) = \epsilon_n \phi^\circ_n({\bf r})\ ,
\end{equation}
i.e. the wavefunctions $\phi^\circ_n({\bf r})$ are just the Bloch states 
$\phi^\circ_{{\bf k}\nu}({\bf r})$ of our 
semiconductor,
and the energy spectrum $\epsilon_n$ coincides with the carrier band 
structure $\epsilon_{{\bf k}\nu}$.
Therefore, from Eq.~(\prr{eq41}) the desired eigenfunctions result to be 
of the form:
\begin{equation}\label{eq44}
\phi_{{\bf k}\nu}({\bf r},t) = 
\phi^\circ_{{\bf k}\nu}({\bf r},t) e^{{i e\over\hbar c}{\bf A}(t)\cdot{\bf r}}\ .
\end{equation}
Apart from a phase-factor, they coincide with the 
conventional Bloch states of the crystal. The reason can be understood 
as follows: 
Also in the presence of the electric field ${\bf F}$, the Hamiltonian 
in Eq.~(\prr{eq40}) is still invariant under a lattice translation 
corresponding to the crystal potential $V^l$. Thus,
the crystal momentum $\hbar{\bf k}$ is still a ``good'' quantum
number and the band dispersion remains the same: 
$\epsilon = \epsilon_{{\bf k}\nu}$.
Therefore, at each time $t$ our time-dependent eigenstates seem to coincide
(a part from the phase-factor) with the Bloch states of the crystal 
and, at a first glance, it is not clear which is the role played 
by the applied field.

In order to answer this question, let us consider again the general form of
our eigenstates in Eq.~(\prr{eq44}).
At the initial time $t_\circ$ the vector potential 
${\bf A}$ is equal to zero and, therefore, the two basis sets coincide:
$
\phi^{ }_{{\bf k}_\circ\nu}({\bf r},t_\circ) = 
\phi^\circ_{{\bf k}\nu}({\bf r})
$.
(Here, ${\bf k}_\circ$ and ${\bf k}$ denote the carrier wavevectors at time 
$t_\circ$ and $t$, respectively. They are, in principle, independent 
quantities, since they correspond to two different eigenvalue 
problems.)
In other words, the Bloch states $\phi^\circ_{{\bf k}\nu}$ can be 
also regarded as the states $\phi^{ }_{{\bf k}_\circ\nu}$  at the initial time 
$t_\circ$ and vice versa. This allows us to rewrite Eq.~(\prr{eq44}) 
as 
\begin{equation}\label{eq45}
\phi_{{\bf k}\nu}({\bf r},t) = 
\phi_{{\bf k}_\circ\nu}({\bf r},t_\circ) 
e^{{i e\over\hbar c}{\bf A}(t)\cdot{\bf r}}
\end{equation}
or equivalently
\begin{equation}\label{eq46}
\phi^\circ_{{\bf k}\nu}({\bf r}) = 
\phi^\circ_{{\bf k}_\circ\nu}({\bf r}) 
e^{-{i e\over\hbar c}{\bf A}(t)\cdot{\bf r}}\ .
\end{equation}
Moreover, in view of the translational symmetry of the crystal, 
at each time $t$ the Bloch states $\phi^\circ$ 
should obey the Bloch theorem \prc{Kittel63}:
\begin{equation}\label{eq47}
\phi^\circ_{{\bf k}\nu}({\bf r+a}) = \phi^\circ_{{\bf k}\nu}({\bf r}) 
e^{i{\bf k}\cdot{\bf a}}\ ,
\end{equation}
where ${\bf a}$ denotes any periodicity vector of the crystal.
If we now apply the Bloch theorem to both sides of Eq.~(\prr{eq46}), 
i.e. at time $t$ and $t_\circ$, we finally obtain:
\begin{equation}\label{eq48}
{\bf k} = {\bf k}_\circ -{e\over\hbar c}{\bf A}(t)\ .
\end{equation}
This result is quite important: the symmetry properties of the Hamiltonian 
require a precise relationship between the (formally independent)
wavevectors ${\bf k}_\circ$ and ${\bf k}$. More specifically, 
the carrier wavevector results to be a 
function of time (${\bf k} = {\bf k}(t)$), i.e. the instantaneous sets of 
basis states $\{\phi_{{\bf k}(t)\nu}({\bf r},t)\}$ (corresponding to 
different times $t$) are mutually connected through a continuous 
time evolution of the wavevector ${\bf k}$.
We can then answer the previous question saying that within the 
vector-potential gauge the application of a homogeneous field ${\bf F}(t)$ 
induces a simple ``drift'' in ${\bf k}$-space of the crystal Bloch states, 
which are therefore not ``distorted'' by the presence of the field.

From a physical point of view, the above equation describes the continuous 
time evolution of a carrier in ${\bf k}$-space induced by the applied 
electric field. In particular, taking into account the explicit form of the 
vector potential ${\bf A}$ given in Eq.~(\prr{eq38}), we have
\begin{equation}\label{eq49}
{\bf k}(t) = {\bf k}_\circ + {e\over \hbar} \int_{t_\circ}^t {\bf F}(t') dt'\ ,
\end{equation}
from which the acceleration theorem in Eq.~(\prr{eq0}) is recovered:
\begin{equation}\label{eq50}
\dot{\bf k}(t) = {e\over\hbar} {\bf F}(t)\ .
\end{equation}
As pointed out in the introduction, this is usually regarded as a 
``semiclassical'' result, i.e. 
as obtained by applying to a Bloch electron the laws of classical mechanics. 
However, the above analysis shows that this is a rigorous 
quantum-mechanical result: the quantum evolution of a
carrier within a given band $\nu$ under the action of a homogeneous 
electric field is rigorously described by the acceleration theorem. 

We want to stress once again that within the vector-potential gauge 
discussed so far the acceleration theorem is just a result of the symmetry 
properties of the crystal and the time-dependent eigenstates 
in Eq.~(\prr{eq44}) describe the quantum analogue of the 
``semiclassical motion'' of a carrier in 
${\bf k}$-space 
(see Fig.~\prr{fig1}). 

Let us now consider the case of a static field, 
i.e. ${\bf F}(t) = {\bf F}_\circ$, applied parallel to a symmetry axis of 
the crystal. In this case, the corresponding vector potential entering
Eq.~(\prr{eq48}) is a linear function of time, which induces a uniform drift 
of the carriers in ${\bf k}$-space along the field direction:
\begin{equation}\label{eq51}
{\bf k}(t) = {\bf k}_\circ + \dot{\bf k} (t-t_\circ) = 
{e {\bf F}_\circ \over\hbar} (t-t_\circ)\ .
\end{equation}
Since the carrier energy ---given by the 
eigenvalue in Eq.~(\prr{eq40})--- coincides with the crystal band structure 
$\epsilon_{{\bf k}\nu}$, its time evolution is:
\begin{equation}\label{eq52}
\epsilon_\nu\left(t\right) = 
\epsilon_\nu\left({\bf k}(t)\right) =
\epsilon_\nu\left({\bf k}_\circ+\dot{\bf k}(t-t_\circ)\right)\ .
\end{equation}
Due to the periodic nature of the band structure as a function of ${\bf k}$, 
i.e. $\epsilon_\nu({\bf k}) = \epsilon_\nu({\bf k+G})$ (${\bf G}$ being a 
reciprocal-lattice vector), the carrier will execute a 
periodic motion in time with a period 
\begin{equation}\label{eq53}
\tau_B = {h\over e F_\circ d}\ ,
\end{equation}
where $d$ is the lattice periodicity (in real space) along the field 
direction.
This coincides with the Bloch period previously introduced (see 
Sect.~\prr{s:int}). 
It corresponds to the time needed for 
the electron to travel from any point $k$ to the energetically equivalent 
point $k+{2\pi\over d}$.

These periodic oscillations of the carrier over the crystal band structure 
(see Fig.~\prr{fig1})
are known as Bloch oscillations. As pointed out in 
Sect.~\prr{s:int}, they were first introduced by Bloch \prc{Bloch28} on the 
basis of semiclassical arguments. However, as for the case of the 
acceleration theorem discussed above, this is a rigorous quantum-mechanical 
result of the vector-potential picture discussed so far.
Such a clear physical 
interpretation of the quantum-mechanical theory in terms of a 
semiclassical picture
is hard to obtain within the scalar-potential gauge presented in the 
following section.

Both the acceleration theorem (\prr{eq50}) and the Bloch-oscillation dynamics 
previously discussed are induced by the noninteracting-carrier Hamiltonian in 
Eq.~(\prr{eq40}) through its time-dependent eigenstates 
$\phi_{{\bf k}(t)\nu}$. Therefore, the Bloch-oscillation dynamics considered 
so far does not account for many-body effects (carrier-carrier and 
carrier-phonon interactions) as well as for the effects induced by the 
time variation of our basis states.
For a more ``realistic'' description of the carrier dynamics within our 
vector-potential picture 
we are then forced to employ the general 
kinetic theory presented in Sect.~\prr{ss:kd}.

As discussed in chapter 6, for the case of a homogeneous semiconductor 
crystal the only relevant terms of 
the single-particle density matrix in our ${\bf k}\nu$ representation are 
those diagonal in ${\bf k}$. 
This property, which is due again to the translational symmetry of the 
Hamiltonian, reduces the set of kinetic variables in 
Eqs.~(\prr{eq24}-\prr{eq25}) 
to the intraband density-matrix elements
\begin{equation}\label{eq54}
f^e_{{\bf k},\alpha\alpha'} = \left\langle c^\dagger_{{\bf k}\alpha} 
c^{ }_{{\bf k}\alpha'} \right\rangle\ , \qquad
f^h_{{\bf -k},\beta\beta'} = \left\langle d^\dagger_{{\bf -k}\beta} 
d^{ }_{{\bf -k}\beta'} \right\rangle
\end{equation}
plus the interband density-matrix elements
\begin{equation}\label{eq55}
p^{ }_{{\bf k},\beta\alpha} = \left\langle d^{ }_{{\bf -k}\beta} 
c^{ }_{{\bf k}\alpha} \right\rangle\ .
\end{equation}
Here, the standard electron-hole picture introduced in Sect.~\prr{ss:ps} has 
been applied to our set of time-dependent eigenstates 
$\phi_{{\bf k}(t)\nu}$: 
the band index $\nu$ (which refers to both 
conduction and valence states) is replaced by two separate band indices 
$\alpha$ and $\beta$ for electrons and holes, respectively, while, due to 
the charge-conjugation symmetry, the hole states are still labeled in terms
of the corresponding valence-electron states, i.e. ${\bf k}^h\beta \equiv 
-{\bf k}^e\beta$.  

Let us now discuss the explicit form of the kinetic equations (\prr{eq30}) 
in our vector-potential picture:
\begin{eqnarray}\label{eq56}
\frac{d}{dt} f^e_{{\bf k},\alpha_1\alpha_2} = \apt 
\frac{d}{dt} f^e_{{\bf k},\alpha_1\alpha_2}\Biggl|_{\bf H} + 
\frac{d}{dt} f^e_{{\bf k},\alpha_1\alpha_2}\Biggl|_\phi \nonumber \\[1ex]
\frac{d}{dt} f^h_{{\bf -k},\beta_1\beta_2} = \apt 
\frac{d}{dt} f^h_{{\bf -k},\beta_1\beta_2}\Biggl|_{\bf H} + 
\frac{d}{dt} f^h_{{\bf -k},\beta_1\beta_2}\Biggl|_\phi \nonumber \\[1ex]
\frac{d}{dt} p^{ }_{{\bf k},\beta_1\alpha_1} = \apt 
\frac{d}{dt} p^{ }_{{\bf k},\beta_1\alpha_1}\Biggl|_{\bf H} + 
\frac{d}{dt} p^{ }_{{\bf k},\beta_1\alpha_1}\Biggl|_\phi \ .
\end{eqnarray}
Here, the first term is induced 
by the system Hamiltonian ${\bf H} = {\bf H}_{sp} + {\bf H}_{mb}$ 
(which does not account for the time 
variation of the basis states) while the second one is induced by the 
time evolution of the 
basis functions $\phi$.

The contributions to the carrier dynamics due to the single-particle 
Hamiltonian ${\bf H}_{sp}$ are given in Eq.~(\prr{eq33}). 
They consist in a ``free rotation'' plus a term 
due to the interaction with the external laser field.
For the case of an ultrafast laser excitation, after the initial carrier 
photogeneration the only non-vanishing 
contributions in Eq.~(\prr{eq33}) are such free-rotation terms.
If we now consider that in our vector-potential representation the energy 
matrix $\epsilon$ in Eq.~(\prr{eq18}) is diagonal,
\begin{equation}\label{eq57}
\epsilon^{e/h}_{{\bf k}l,{\bf k}'l'} = \epsilon^{e/h}_{{\bf k}l} 
\delta_{{\bf kk}'} \delta_{ll'}\ ,
\end{equation}
the single-particle contributions after the initial 
photoexcitation reduce to:
\begin{eqnarray}\label{eq58}
\frac{d}{dt} f^e_{{\bf k},\alpha_1\alpha_2}\Biggl|_{sp} = \apt \frac{1}{i\hbar} 
\left[\epsilon^e_{{\bf k}\alpha_2} -
\epsilon^e_{{\bf k}\alpha_1}\right] f^e_{{\bf k},\alpha_1\alpha_2} 
\nonumber \\
\frac{d}{dt} f^h_{{\bf -k},\beta_1\beta_2}\Biggl|_{sp} = \apt 
\frac{1}{i\hbar} \left[\epsilon^h_{{\bf -k}\beta_2} -
\epsilon^h_{{\bf -k}\beta_1}\right] f^h_{{\bf -k},\beta_1\beta_2} \nonumber 
\\[1ex]
\frac{d}{dt} p^{ }_{{\bf k},\beta_1\alpha_1}\Biggl|_{sp} = \apt 
\frac{1}{i\hbar} 
\left[\epsilon^h_{{\bf -k}\beta_1} + \epsilon^e_{{\bf k}\alpha_1}\right] 
p^{ }_{{\bf k},\beta_1\alpha_1} \ .
\end{eqnarray}
As we can see, the above equations describe a set of independent many-level 
systems, i.e. one for each ${\bf k}$ value. In addition, there is no 
coupling between different kinetic variables.
If, in particular, we assume a diagonal initial condition 
\begin{equation}\label{eq59}
f^{e/h}_{{\bf k},ll'} \equiv f^{e/h}_{{\bf k},l} \delta_{ll'}\ , \qquad
p^{ }_{{\bf k},ll'} \equiv p^{ }_{{\bf k},l} \delta_{ll'}
\end{equation}
(which is well fulfilled for the case of a laser photoexcitation of 
standard bulk semiconductors as well as 
superlattice structures), 
the kinetic equations (\prr{eq58}) for the non-zero density-matrix elements, 
i.e. for the diagonal ones, reduce to:
\begin{equation}\label{eq60}
\frac{d}{dt} f^{e/h}_{\pm{\bf k},l} = 0\ , \qquad
\frac{d}{dt} p^{ }_{{\bf k},l} = 0\ , 
\end{equation}
where the $\pm$ sign refers, respectively, to electrons and holes.
If we now remember that in our vector-potential representation the 
wavevector ${\bf k}$ is itself a function of time (see Eq.~(\prr{eq48})), 
the above kinetic equations can be rewritten as:
\begin{equation}\label{eq61}
\frac{\partial}{\partial t} f^{e/h}_{\pm{\bf k},l} 
\pm \dot{\bf k}\cdot\nabla_{\bf k} f^{e/h}_{\pm{\bf k},l} = 0\ , \qquad
\frac{\partial}{\partial t} p^{ }_{{\bf k},l} 
+ \dot{\bf k}\cdot\nabla_{\bf k} p^{ }_{{\bf k},l} = 0\ .
\end{equation}
For both distribution functions $f$ and interband 
polarizations $p$ we obtain a simple drift equation whose general solution 
is of the form:
\begin{equation}\label{eq62}
y\left({\bf k}(t),t\right) = y\left({\bf k}_\circ,t_\circ\right)\ ,
\end{equation}
i.e. the function at time $t$ is obtained through a rigid shift 
$\Delta{\bf k} = {\bf k}-{\bf k}_\circ$ of the 
function at the initial time $t_\circ$. 
Such drift in ${\bf k}$-space, induced by the external field ${\bf F}$, 
is again described by the acceleration theorem in 
Eq.~(\prr{eq50}).
Therefore, as expected, the carrier dynamics described by the above kinetic
equations (which accounts for the single-particle Hamiltonian only)  
is fully equivalent to the Bloch-oscillation dynamics discussed above.
However, such relatively simple picture of the carrier motion does not 
account for the time-dependence of our basis states
as well as for many-body effects, e.g. carrier-carrier and 
carrier-phonon interactions. 

The contributions to the carrier dynamics induced by the time variation of 
the basis states are given in Eq.~(\prr{eq31}).
In our vector-potential representation (\prr{eq44}) the explicit form of 
the matrix elements $Z^{e/h}_{ll'}$ introduced in Eq.~(\prr{eq29}) is
\begin{equation}\label{eq63}
Z^{e/h}_{{\bf k}l,{\bf k}'l'} = Z^{e/h}_{{\bf k},ll'} \delta_{{\bf kk}'} 
\end{equation}
with
\begin{equation}\label{eq64}
Z^{e/h}_{{\bf k},ll'} = \pm e 
\left(\delta_{ll'}-1\right)
\int d{\bf r} 
\phi^{e/h *}_{{\bf k}l} \left[{\bf F}(t)\cdot{\bf r}\right]
\phi^{e/h}_{{\bf k}l'} \ .
\end{equation}
They result to be strictly related to the matrix elements of the scalar 
potential in Eq.~(\prr{eq39}), as we will discuss in the following section. 
The ${\bf k}l \to {\bf k}'l'$ transitions 
induced by the time variation of the basis states are always diagonal in 
${\bf k}$; this reflects the momentum conservation in the carrier-field 
interaction, i.e. since the momentum ${\bf q}$ corresponding to a 
space-independent field is equal to zero, the initial and final carrier 
wavevectors coincide.
Moreover, there are no intraband ($l = l'$) transitions,
which confirms that the action of the field within a given band is fully 
described by the drift terms in Eq.~(\prr{eq61}).

If we now rewrite Eq.~(\prr{eq31}) in our vector-potential representation 
taking into account the explicit form of the above matrix elements, we 
finally obtain:
\begin{eqnarray}\label{eq65}
\frac{d}{dt} f^e_{{\bf k},\alpha_1\alpha_2}\Biggl|_\phi = \apt \frac{1}{i\hbar} 
\sum_{\alpha_3\alpha_4}
\left[
Z^e_{{\bf k},\alpha_2\alpha_4} \delta_{\alpha_1\alpha_3} -
Z^e_{{\bf k},\alpha_3\alpha_1}\delta_{\alpha_2\alpha_4}\right] 
f^e_{{\bf k},\alpha_3\alpha_4} 
\nonumber \\
\frac{d}{dt} f^h_{{\bf -k},\beta_1\beta_2}\Biggl|_\phi = \apt 
\frac{1}{i\hbar} 
\sum_{\beta_3\beta_4}
\left[Z^h_{{\bf -k},\beta_2\beta_4}\delta_{\beta_1\beta_3} -
Z^h_{{\bf -k},\beta_3\beta_1}\delta_{\beta_2\beta_4}\right] 
f^h_{{\bf -k},\beta_3\beta_4} \nonumber 
\\[1ex]
\frac{d}{dt} p^{ }_{{\bf k},\beta_1\alpha_1}\Biggl|_\phi = \apt 
\frac{1}{i\hbar} 
\sum_{\alpha_2\beta_2}
\left[Z^h_{{\bf -k},\beta_1\beta_2}\delta_{\alpha_1\alpha_2} + 
Z^e_{{\bf k}\alpha_1\alpha_2}\delta_{\beta_1\beta_2}\right] 
p^{ }_{{\bf k},\beta_2\alpha_2} \ .
\end{eqnarray}
From the above kinetic equations we clearly see that the time variation of 
our basis states $\phi$ induces ``vertical'' (i.e. ${\bf k} = {\bf k}'$) 
transitions between different bands. Such interband coupling is 
the well known Zener tunneling \prc{Zener34}. 
This effect is usually described
as an interband transition induced by the scalar potential 
$-{\bf F}\cdot{\bf r}$, which is also evident from the explicit 
form of the Zener matrix elements in Eq.~(\prr{eq64}). 
However, within our vector-potential picture the Zener tunneling 
originates from the time variation of our accelerated Bloch states in 
Eq.~(\prr{eq44}).

As discussed in Ref.~\prc{Krieger86}, 
for the case of bulk semiconductors this interband 
coupling results to be very limited even for the case of high applied 
fields. Therefore, the Bloch-oscillation scenario of the semiclassical 
theory (see Fig.~\prr{fig1}) 
is practically unmodified by Zener tunneling.
For the case of interest, i.e. that of semiconductor superlattices, 
interminiband Zener tunneling is expected to play a significant role in the 
high-field regime. 
However, for relatively low fields (up to $10^4$V/cm) the effect 
is again negligible and the Bloch-oscillation regime is fully recovered.
In this case, the time-scale of Zener-tunneling processes is much longer than 
the Bloch-oscillation period $\tau_B$. 
Therefore, the effect due to the time variation of our 
basis states is negligible, i.e. the time variation can be regarded as an 
``adiabatic transformation''.

Let us finally consider the role played by many-body effects, i.e. 
carrier-carrier and carrier-phonon interactions.
As discussed in Sect.~\prr{ss:kd} as well as in chapter 6, 
these many-body effects can be divided into coherent and incoherent 
contributions.
With coherent contributions we refer to first-order terms in the many-body 
Hamiltonian ${\bf H}_{mb}$. 
Since we neglect coherent-phonon states, 
the only non-zero contributions originate from carrier-carrier interaction.
The explicit form of these Hartree-Fock terms in our vector-potential 
representation is obtained from Eqs.~(\prr{eq34}-\prr{eq36}) by replacing 
the generic labels $i$ and $j$ with ${\bf k}\alpha$ and ${\bf -k}\beta$, 
respectively.

From a physical point of view, these coherent contributions give rise to 
excitonic and band-renormalization effects, which result in a modification 
of  the single-particle energy spectrum. 
In our case, these excitonic effects may lead to modifications of the 
Bloch-oscillation dynamics, e.g. small variations of the Bloch period. 
As we will see, within the Wannier-Stark picture such excitonic effects 
manifest themselves in a modification of the single-particle Wannier-Stark 
energy levels, which are not equidistantly spaced anymore \prc{Dignam90}.

Let us now come to the role played by incoherent contributions, i.e. 
second-order contributions in the many-body Hamiltonian.
As pointed out in Sect.~\prr{ss:kd}, these terms are usually 
treated within the usual Markov approximation.
The resulting contributions describe, in general, second-order transitions 
connecting all possible kinetic variables, i.e. all possible density-matrix
elements. 
The transitions connecting diagonal density-matrix elements, i.e. 
distribution functions, can be easily described in terms of 
stochastic scattering processes, i.e. due to the scattering with a partner 
carrier or with a phonon, the electron may undergo a transition from an 
initial state ${\bf k}\nu$ to a final state ${\bf k}'\nu'$.
On the contrary, for the transitions involving non-diagonal terms the 
second-order coupling is not positive-definite, i.e. it is not a rate, 
and the intuitive scattering picture cannot be employed.

It is not in
the spirit of this chapter to derive and discuss the explicit form of 
the second-order carrier-carrier and carrier-phonon contributions.
From a physical point of view, both carrier-carrier and carrier-phonon 
scattering processes give rise to energy relaxation and dephasing. 
It is well known that, due to scattering events, 
a photogenerated carrier distribution will relax both energy and momentum 
and, in addition, it will lose its internal degree of coherence, which 
corresponds to a decay of the interband polarizations.
This stochastic dynamics may strongly influence the de\-ter\-mi\-ni\-stic 
Bloch-oscilla\-tion regime discussed so 
far \prc{vonPlessen94,Rossi95a,Rossi95b}.

As we will see from some simulated experiments reported in 
Sect.~\prr{s:sse}, the role played by carrier-carrier and 
carrier-phonon scattering strongly depends on the physical conditions 
considered, e.g. carrier density, excitation energy, and lattice temperature.
When the scattering rate corresponding to the dominant interaction 
mechanism is much larger than the Bloch oscillation frequency $\omega_B = 
{2\pi\over\tau_B}$, the Bloch oscillations are not suppressed, 
i.e. the carriers perform on average several Bloch oscillations between two
scattering events.
On the contrary, if the scattering rate is larger than the Bloch 
frequency, the carrier cannot execute a full oscillation 
without scattering. In this case, the Bloch oscillations are totally 
suppressed and we deal with a diffusive-transport regime.

As discussed in the introductory part of this chapter, in bulk 
semiconductors also for very high fields the Bloch-oscillation period 
is larger than the typical scattering times. 
On the contrary, in semiconductor superlattices the Bloch period is at 
least one order of magnitude smaller than in bulk systems and, therefore, 
comparable or even smaller than the typical scattering times.

This allows us to answer the controversial question: ``do Bloch oscillations
really exist?".
The analysis presented in this section shows that, in the absence of 
scattering events, Bloch oscillations exist and, contrary to the early 
papers by Rabinovitch and Zak \prc{Rabinovitch72}, 
they are not significantly affected by Zener tunneling, 
both for bulk and superlattices.

On the contrary, due to scattering events, Bloch oscillations are fully 
suppressed in bulk semiconductors but they still survive in superlattices, 
as confirmed by several 
experiments \prc{Feldmann92,Leo92,Waschke93,Roskos94,vonPlessen94}.

In the following section we will discuss the so-called Wannier-Stark 
picture. Contrary to the vector-potential approach discussed so far, this 
will correspond to a scalar-potential representation.
In particular, we will study the link between the two pictures showing 
that phenomena which are peculiar of one picture can be equally described 
within the second one. 
\subsection{The Wannier-Stark picture}\label{ss:wsp}

Within the scalar-potential gauge (\prr{eq39}), the eigenvalue 
equation (\prr{eq37}) reduces to:
\begin{equation}\label{eq66}
H^\circ_c \phi({\bf r}) = \left[
-{\hbar^2\nabla^2_{\bf r}\over 2 m_\circ} + V^l({\bf r}) -e{\bf 
F}\cdot{\bf r}
\right] \phi({\bf r}) = \epsilon \phi({\bf r})\ .
\end{equation}
In this gauge the scalar potential is space dependent but, for the case of 
a static field considered in this section, it is always 
time independent. 

As originally pointed out by Wannier \prc{Wannier60}, 
due to the translational symmetry of the crystal potential 
$V^l({\bf r}) = V^l({\bf r}+{\bf d})$, ${\bf d}$ being the primitive 
lattice vector along the field direction, if $\phi({\bf r})$ is an 
eigenfunction of the Hamiltonian in Eq.~(\prr{eq66}) with eigenvalue 
$\epsilon$, 
then any $\phi({\bf r}+n{\bf d})$ is also an
eigenfunction with eigenvalue 
$\epsilon+n \Delta\epsilon$, where $\Delta\epsilon = e F d$ is the 
so-called Wannier-Stark splitting.
However, as pointed out by Zak \prc{Zak68a},
for the case of an infinite crystal the scalar potential 
$-{\bf F}\cdot{\bf r}$ is not bounded, which implies a continuous energy 
spectrum.
This was the starting point of the long-standing controversy on the existence 
of Wannier-Stark ladders discussed in Sect.~\prr{s:hb}. 

Today, after three decades, we know that the problem is by itself 
ill-defined and eigenstates can be defined only 
asymptotically \prc{Nenciu91}. 
In particular, there exist  
ladders of metastable Wannier-Stark states weakly coupled (through Zener 
tunneling) to a continuous 
energy spectrum.

We will now study the explicit form of these Wannier-Stark states.
As a starting point, let us write the generic eigenstate $\phi$ 
as a superposition of conventional Bloch states $\phi^\circ_{{\bf k}\nu}$, 
i.e. let us move to the so-called crystal-momentum representation (CMR):
\begin{equation}\label{eq67}
\phi({\bf r}) = 
\sum_{{\bf k}\nu} s_{{\bf k}\nu} \phi^\circ_{{\bf k}\nu}({\bf r}) =
\sum_{{\bf G}\nu} {1\over\Omega}
\int^\Omega d{\bf k} s_{{\bf k+G}\,\nu} \phi^\circ_{{\bf 
k+G}\,\nu}({\bf r})
\ ,
\end{equation}
where ${\bf G}$ is a generic reciprocal-lattice vector while $\Omega$ 
denotes the volume of the first Brillouin zone. 
Due to the periodicity of the Bloch 
states in ${\bf k}$-space 
($\phi^\circ_{{\bf k+G}\,\nu} = \phi^\circ_{{\bf k}\nu}$), 
we are allowed to limit the above 
expansion to the first Brillouin zone, which 
implies to impose the same periodicity on the coefficients 
$s_{{\bf k}\nu}$: $s_{{\bf k+G}\,\nu} = s_{{\bf k}\nu}$.
By inserting the above expansion (limited to the first Brillouin zone) in 
Eq.~(\prr{eq66}), our eigenvalue problem 
can be rewritten as:
\begin{equation}\label{eq68}
\sum_{\nu'} {1\over\Omega} 
\int^\Omega d{\bf k}' 
H^\circ_{{\bf k}\nu,{\bf k}'\nu'} s_{{\bf k}'\nu'} = 
\epsilon s_{{\bf k}\nu}\ ,
\end{equation}
where 
\begin{equation}\label{eq69}
H^\circ_{{\bf k}\nu,{\bf k}'\nu'} = 
\epsilon_{{\bf k}\nu} \delta_{{\bf k}\nu,{\bf k}'\nu'} +
H^{\bf F}_{{\bf k}\nu,{\bf k}'\nu'}
\end{equation}
are the matrix elements of the 
Hamiltonian $H^\circ_c$ in the CMR. The first term corresponds to the 
perfect (field-free) crystal while the second one describes the 
scalar-potential term: 
\begin{equation}\label{eq70}
H^{\bf F}_{{\bf k}\nu,{\bf k}'\nu'} = -e \int d{\bf r} 
\phi^{\circ *}_{{\bf k}\nu} \left[{\bf F}\cdot{\bf r}\right]
\phi^\circ_{{\bf k}'\nu'} \ .
\end{equation}
Following the approach discussed in Ref.~\prc{Quade94}, the above 
scalar-potential matrix elements can be divided into intraband ($\nu = 
\nu'$) and interband ($\nu \ne \nu'$) terms:
The intraband terms can always be written as a drift 
operator \prc{Quade94,Meier94,Meier95a}
\begin{equation}\label{eq71}
{1\over\Omega_\parallel} \int^{\Omega_\parallel} dk'_\parallel 
H^{\bf F}_{{\bf k}\nu,{\bf k}'\nu} = -i e F 
\delta_{k_\perp k'_\perp} {\partial\over\partial k_\parallel}
\end{equation}
($k_\parallel$ and $k_\perp$ being,respectively, the components of the 
wavevector ${\bf k}$ parallel and perpendicular to the field)
while the non-diagonal terms coincide with the Zener-tunneling matrix 
elements $Z$ in Eq.~(\prr{eq63}). 
As already pointed out in Sect.~\prr{ss:bop}, the Zener 
tunneling, which in the vector-potential representation is induced by the 
time variation of the basis states, corresponds to interband transitions 
induced by the scalar potential in Eq.~(\prr{eq39}).

For moderate values of the applied field, Zener tunneling to other bands 
can be neglected 
and, by inserting Eqs.~(\prr{eq69},\prr{eq71}) into 
Eq.~(\prr{eq68}), 
our eigenvalue equation reduces to
\begin{equation}\label{eq72}
{\partial\over\partial k_\parallel} s_{{\bf k}\nu} = 
-{i\over e F} \left(\epsilon_{{\bf k}\nu} - \epsilon\right) s_{{\bf k}\nu}
\ ,
\end{equation}
whose solution is given by:
\begin{equation}\label{eq73}
s_{{\bf k}\nu} = \delta_{k_\perp \tilde{k}_\perp} 
e^{-{i\over eF} \int_0^{k_\parallel} 
(\epsilon_{k'_\parallel \tilde{k}_\perp\nu}-\epsilon) 
dk'_\parallel}\ .
\end{equation}
As expected, the coefficients are diagonal with respect to $k_\perp$, i.e. 
the linear combination in Eq.~(\prr{eq67}) will only involve Bloch states 
with the same perpendicular component $k_\perp$. This reflects the 
translational symmetry of the Hamiltonian with respect to the plane 
perpendicular to the field.
Moreover, 
the periodicity condition
$s_{{\bf k+G}\,\nu} = s_{{\bf k}\nu}$ (applied along the field direction) 
requires that
\begin{equation}\label{eq74}
{1\over eF} \int_{-{\pi\over d}}^{+{\pi\over d}} 
\left(\epsilon_{k_\parallel k_\perp\nu}-\epsilon\right) dk_\parallel 
= 2\pi n
\end{equation}
which, in turn, tells us that the only allowed energy values are
\begin{equation}\label{eq75}
\epsilon = \epsilon^{k_\perp n\nu} = \epsilon^{k_\perp 0\nu} + n\,eFd
\end{equation}
with
\begin{equation}\label{eq76}
\epsilon^{k_\perp 0\nu} = {d\over 2\pi} \int_{-{\pi\over d}}^{+{\pi\over d}} 
\epsilon_{k_\parallel k_\perp\nu} dk_\parallel\ .
\end{equation}
What we obtain is the Wannier-Stark ladder mentioned above, whose central 
($n = 0$) value is given by the average of the band 
along the field direction (for a given $k_\perp$).

Let us now discuss the corresponding wavefunctions.
By inserting in the expansion (\prr{eq67}) 
the explicit form of the coefficients 
$s_{{\bf k}\nu}$ given in Eq.~(\prr{eq73}) we have:
\begin{equation}\label{eq77}
\phi^{k_\perp n\nu}({\bf r}) = {d\over 2\pi} 
\int_{-{\pi\over d}}^{+{\pi\over d}} dk_\parallel
e^{-{i\over eF} \int_0^{k_\parallel} (\epsilon_{k'_\parallel k_\perp\nu}-
\epsilon^{k_\perp n\nu}) 
dk'_\parallel}
\phi^\circ_{k_\parallel k_\perp\nu}({\bf r})\ .
\end{equation}
They are the so-called Wannier-Stark states shown in Fig.~\prr{fig2}.
For each band $\nu$ (and for any given $k_\perp$), 
we have a set of energetically equidistant states
$\phi^{k_\perp n\nu}$, obtained from the central ($n = 0$) 
state $\phi^{k_\perp 0\nu}$ 
through the spatial translation 
${\bf r} \to {\bf r}+n{\bf d}$ 
discussed above. 
As schematically depicted in Fig.~\prr{fig2}, each state is localized around 
one of the atomic cells of the superlattice and the degree of localization 
depends on the 
strength of the applied field $F$. 
More precisely, when the Wannier-Stark energy $eFd$ is much smaller than 
the superlattice miniband width, the wavefunction $\phi^{k_\perp n\nu}$ 
are weakly localized, they extend over several elementary cells. 
On the contrary, when $eFd$ is comparable or larger than the miniband width, 
the localization increases and the function results to be significantly 
different from zero only in one cell.
Since the miniband width of the holes is smaller than that of the 
electrons, 
the hole states exhibit a stronger Wannier-Stark localization 
(see Fig.~\prr{fig2}).

Due to the neglect of interband Zener tunneling, each Wannier-Stark state 
$\phi^{k_\perp n\nu}$ in Eq.~(\prr{eq77}) is obtained as a linear combination 
of Bloch states belonging to the same band $\nu$. 
Moreover, we see that all 
Bloch states have the same weight in the expansion, i.e. the coefficients 
are just phase-factors.

The coefficients $s$ introduced so far can be regarded as the matrix 
elements of a unitary transformation connecting the Bloch to 
the Wannier-Stark representation. It is then clear that the inverse 
transformation allows us to write any Bloch state as a linear combination 
of Wannier-Stark states:
\begin{equation}\label{eq78}
\phi^\circ_{{\bf k}\nu} = \sum_n s^{n *}_{{\bf k}\nu} 
\phi^{k_\perp n\nu}({\bf r})\ .
\end{equation}

The Wannier-Stark states in Eq.~(\prr{eq77}) can now be used as 
basis states for our kinetic description.
The kinetic variables in the Wannier-Stark representation will be 
formally the same as for the vector-potential picture discussed in 
Sect.~\prr{ss:bop}.
They are defined according to Eqs.~(\prr{eq54},\prr{eq55}) 
where the three-dimensional 
wavevector ${\bf k}$ is replaced by its perpendicular component $k_\perp$ 
while the band index $\alpha/\beta$ is replaced by the same band index plus 
the Wannier-Stark ladder index $n$.

Provided the above label substitution, the time evolution of the new kinetic 
variables is again described by Eq.~(\prr{eq56}).
However, within our Wannier-Stark representation the basis states are 
time-independent and, therefore, the $\phi$-contributions in 
Eq.~(\prr{eq56}) are equal to zero.
Moreover, contrary to the vector-potential case, the energy matrix 
$\epsilon$ in Eq.~(\prr{eq18}) is not diagonal. 
The diagonal terms are now given by the Wannier-Stark ladders in 
Eq.~(\prr{eq75}) while the non-diagonal terms 
are once again Zener-tunneling matrix elements between different 
Wannier-Stark states.

As already pointed out in Sect.~\prr{ss:kd}, the coherent contributions 
entering our kinetic equations are independent from 
the quantum-mechanical representation considered. This tells us that the 
Zener-tunneling contributions to the equations of motion in the 
scalar- and vector-potential representations should coincide.
In fact, in spite of their different physical interpretations 
(in the vector-potential 
picture they are induced by the time variation of the basis states while in 
the scalar-potential one they are due to interband transitions induced by the 
field Hamiltonian), their formal structure is exactly the same, 
as can be seen by comparing 
Eqs.~(\prr{eq31}) and (\prr{eq33}).

On the contrary, incoherent contributions, i.e. scattering terms derived 
within Markov approximation, are expected to be representation-de\-pen\-dent.
In particular, within the vector-potential picture the scattering terms are 
usually derived by neglecting the time dependence of the basis states 
$\phi_{{\bf k}(t)\nu}$, 
which corresponds to neglecting the action of the field, i.e. the 
time variation of the carrier wavevector ${\bf k}(t)$,  
during the collision.
Within the Wan\-nier-Stark picture, the basis states are time-independent and 
the standard Markov limit automatically accounts for the so-called 
intracollisional field effect \prc{Quade94,Brunetti89,Rossi92}.
However, for moderate values of the applied field incoherent 
contributions evaluated in the scalar- and vector-potential representations 
coincide.

Before concluding this section, let us try to better understand the physical 
link between 
the accelerated Bloch states of the vector-potential picture and the 
Wannier-Stark states of the scalar-potential representation.
Both basis sets have been introduced as eigenstates of two equivalent 
Hamiltonians, corresponding to the two different electromagnetic gauges 
(see Eqs.~(\prr{eq40}) and (\prr{eq66})).
However, it is well known \prc{Schiff55} that the solutions of the 
corresponding time-dependent Schr\"odinger equations coincide (a part from 
a phase-factor which is physically irrelevant).
More specifically, let us consider the generic time-dependent Schr\"odinger 
equation 
\begin{equation}\label{eq79}
i\hbar {d \over dt} \psi({\bf r},t) = H(t) \psi({\bf r},t)
\end{equation}
together with the corresponding eigenvalue problem
\begin{equation}\label{eq80}
H(t) \phi_\lambda({\bf r},t) = \epsilon_\lambda(t) \phi_\lambda({\bf r},t)\ .
\end{equation}
The generic solution $\psi$ at time $t$ is given by a linear combination of
the eigenstates $\phi$ according to the time evolution induced by the 
Hamiltonian $H$:
\begin{equation}\label{eq81}
\psi({\bf r},t) = \sum_\lambda S_\lambda 
e^{-{i\over\hbar}\int_{t_\circ}^t \epsilon_\lambda(t') dt'}
\phi_\lambda({\bf r},t)\ ,
\end{equation}
which, for the case of the vector-potential Hamiltonian (\prr{eq40}) and 
the corresponding eigenstates in Eq.~(\prr{eq44}) 
reduces to:
\begin{equation}\label{eq82}
\psi({\bf r},t) = e^{{i e\over\hbar c} {\bf A}(t)\cdot{\bf r}}
\sum_{{\bf k}_\circ\nu} S_{{\bf k}_\circ\nu}  
e^{-{i\over\hbar}\int_{t_\circ}^t \epsilon_{{\bf k}(t')\nu} dt'}
\phi^\circ_{{\bf k}(t)\nu}({\bf r})\ ,
\end{equation}
where ${\bf k}_\circ = {\bf k}(t_\circ)$ denotes the carrier wavevector 
at the initial time $t_\circ$.
As discussed in Ref.~\prc{Krieger86}, it is always possible to consider 
the gauge transformation connecting the above vector-potential picture 
to the scalar-potential one. The generator of this gauge 
transformation is the function $-{\bf A}(t)\cdot{\bf r}$,
which tells us that, going from the vector- to the scalar-potential 
picture,
the first phase-factor in Eq.~(\prr{eq82}) cancels exactly with the 
corresponding phase-factor of the gauge transformation.
Thus, the time-dependent wave function (\prr{eq82}) in the scalar-potential 
gauge reduces to:
\begin{equation}\label{eq83}
\psi({\bf r},t) = 
\sum_{{\bf k}_\circ\nu} S_{{\bf k}_\circ\nu}  
e^{-{i\over\hbar}\int_{t_\circ}^t \epsilon_{{\bf k}(t')\nu} dt'}
\phi^\circ_{{\bf k}(t)\nu}({\bf r})\ .
\end{equation}
This is a linear combination of the so-called Houston 
states \prc{Houston40} originally introduced as time-dependent solutions 
of the scalar-potential Schr\"odinger equation.

On the other hand, within the Wannier-Stark representation (\prr{eq77}), the
linear combination in Eq.~(\prr{eq81}) reduces to:
\begin{equation}\label{eq84}
\psi({\bf r},t) = \sum_{k_\perp n\nu} S^{k_\perp n\nu} 
e^{-{i\over\hbar} \epsilon^{k_\perp n\nu} (t-t_\circ)}
\phi^{k_\perp n\nu}({\bf r})\ .
\end{equation}
It is then clear that for a given initial condition $\psi({\bf r},t_\circ)$
the two last linear combinations must give at any time $t$ the same 
wavefunction $\psi$.
Let us consider as initial condition a single Wannier-Stark state 
$\phi^{k_\perp n\nu}$. According to Eq.~(\prr{eq84}), at time $t$ the 
function $\psi$ differs from that at time $t_\circ$ only by a 
phase-factor, which implies that this will be a stationary state, i.e. the 
wavefunction will remain localized around a given cell and $|\psi|^2$ will 
not change in time.
Therefore, it should be possible to choose the coefficients $S$ 
entering Eq.~(\prr{eq83}) in such a way that the corresponding expansion in 
terms of Houston states will provide the same stationary Wannier-Stark 
state. 

In order to determine the explicit form of the coefficients corresponding 
to a stationary state, let us rewrite 
Eq.~(\prr{eq83}), replacing at each time $t$ the sum over ${\bf k}_\circ$ 
with an equivalent sum over the instantaneous 
${\bf k} = {\bf k}_\circ+\dot{\bf k} 
(t-t_\circ)$ (see Eq.~(\prr{eq51})):
\begin{equation}\label{eq85}
\psi({\bf r},t) = 
\sum_{{\bf k}\nu} S_{{\bf k}-\dot{\bf k}(t-t_\circ)\nu}  
e^{-{i\over\hbar}\int_{t_\circ}^t \epsilon_{{\bf k}(t')\nu} dt'}
\phi^\circ_{{\bf k}\nu}({\bf r})\ .
\end{equation}
The stationary-state condition corresponds to impose that each individual 
term entering the above expansion will 
evolve in time according to the constant Wannier-Stark energy 
$\epsilon^{k_\perp n\nu}$,
\begin{equation}\label{eq86}
S_{{\bf k}-\dot{\bf k}(t-t_\circ)\nu}  
e^{-{i\over\hbar}\int_{t_\circ}^t \epsilon_{{\bf k}(t')\nu} dt'}
\propto e^{-{i\over\hbar} \epsilon^{k_\perp n\nu} (t-t_\circ)}\ ,
\end{equation}
which implies that for each time $t$ 
\begin{equation}\label{eq87}
S_{{\bf k}-\dot{\bf k}(t-t_\circ)\nu}  
\propto
e^{{i\over\hbar}\int_{t_\circ}^t 
\left(\epsilon_{{\bf k}(t')\nu} - \epsilon^{k_\perp n\nu}\right) dt'}\ .
\end{equation}
The above time integral over $t'$ can be translated into a corresponding 
integral over $k' = k_\parallel(t')$:
\begin{equation}\label{eq88}
e^{{ie\over F}\int_{k_\circ}^k 
\left(\epsilon_{k'\nu} - \epsilon^{k_\perp n\nu}\right) dk'}
= e^{{ie\over F}\int_0^k
\left(\epsilon_{k'\nu} - \epsilon^{k_\perp n\nu}
\right) dk'}
e^{-{ie\over F}\int_0^{k_\circ} 
\left(\epsilon_{k'\nu} - \epsilon^{k_\perp n\nu}\right) dk'}\ .
\end{equation}
Since the first phase-factor on the right-hand side is time-independent, we 
finally obtain:
\begin{equation}\label{eq90}
S_{{\bf k}_\circ\nu} \propto 
e^{-{ie\over F}\int_0^{k_\circ} 
\left(\epsilon_{{\bf k}'\nu} - \epsilon^{k_\perp n\nu}\right) dk'}\ .
\end{equation}
As expected, the coefficients $S$ of a stationary state 
coincide with the coefficients $s$ in Eq.~(\prr{eq73}). 
Thus, the linear combinations of accelerated Bloch states corresponding 
to stationary states are just the Wannier-Stark states introduced in 
Eq.~(\prr{eq77}) as eigenstates of the scalar-potential Hamiltonian.

From the above analysis, we see that within a time-dependent approach the 
scalar- and vector-potential pictures are totally equivalent.
According to the initial condition (i.e. depending on the coefficients $S$), we 
may have a Bloch-oscillation as well as a Wannier-Stark scenario, or any 
intermediate regime.
If we consider, for example, the case of a laser excitation 
whose energy spectrum is concentrated around a well defined frequency, this 
will generate a distribution of photoexcited carriers 
strongly peaked about a particular $k$. Each carrier will then be described 
by a single accelerated Bloch state $\phi_{{\bf k}\nu}$ and will execute 
Bloch oscillations. Thus, the overall motion of this packet in ${\bf 
k}$-space will resemble the periodic motion of a single electron 
``prepared'' in a Bloch state (see Fig.~\prr{fig3}).
As we will discuss in the following section, such Bloch-oscillation dynamics 
can be monitored via four-wave-mixing experiments or THz-signal 
measurements.

On the contrary, if we perform a linear-absorption measurement using as a 
light source a laser with uniform spectral distribution, we generate a 
uniform distribution of photoexcited carriers in ${\bf k}$-space with 
their proper phase coherence (described by the corresponding interband
polarizations). Such distribution is shifted in ${\bf k}$-space but, 
being almost constant, there is no macroscopic effect.
This resembles the situation corresponding to a single electron prepared in 
a Wannier-Stark state, which is a uniform superposition of Bloch states. 
This is confirmed by optical-absorption investigations which clearly show the 
Wannier-Stark energy quantization (see Fig.~\prr{fig7}).
\section{Some simulated experiments}\label{s:sse}

In this section, we will review recent simulated experiments of the 
ultrafast carrier dynamics in semiconductor 
superlattices \prc{Rossi95a,Rossi95b,Meier95b,Je95,Koch95,Rossi96}.
They are based on a generalized Monte Carlo 
solution \prc{Kuhn92a,Kuhn92b,Rossi94,Haas96} 
of the set of kinetic 
equations (so-called semiconductor Bloch equations) derived in 
Sect.~\prr{ss:kd}.
This generalized Monte Carlo approach, successfully applied for the 
interpretation of ultrafast coherent phenomena in bulk 
semiconductors 
\prc{Lohner93,Leitenstorfer94a,Leitenstorfer94b,Leitenstorfer96},
is based on a combined solution of our kinetic equations \prc{Haas96}:
the coherent contributions are evaluated by means of a direct numerical
integration while the incoherent ones are ``sampled'' by means of a
conventional Monte Carlo simulation in the three-dimensional 
${\bf k}$-space.

This generalized Monte Carlo method has been recently applied to 
semiconductor superlattices.
As described in Refs.~\prc{Rossi95a,Meier95b},
the simulation sche\-me is based on the Bloch-state representation of the 
vector-potential picture introduced in Sec.~\prr{ss:bop}.

The following superlattice model has been employed:
The energy dispersion and the corresponding 
wavefunctions along the growth direction ($k_\parallel$)
are computed within the well known
Kronig-Penney model \prc{Bastard89}, while 
for the in-plane direction ($k_\perp$) 
an effective-mass model has been used.
Starting from these three-dimensional wavefunctions 
$\phi^\circ_{{\bf k}\nu}$, the 
various carrier-carrier as well as 
carrier-phonon matrix elements are numerically computed (see 
Eqs.~(\prr{eq21}) and (\prr{eq23})).
They are, in general, functions
of the various miniband indices and depend separately 
on $k_\parallel$ and $k_\perp$, 
thus fully reflecting the anisotropic nature  of the 
superlattice structure.

Only coupling to GaAs bulk phonons has been considered.
This, of course, is a simplifying approximation which neglects any
superlattice effect on the phonon dispersion, such as
confinement of optical modes in the wells and in the barriers,
and the presence of interface modes \prc{Molinari94}.
However, while these modifications
have important consequences for phonon spectroscopies
(like Raman scattering), they are far less decisive for transport
phenomena. Indeed, by now it is well known \prc{Molinari94,Ruecker92}
that the total scattering rates are sufficiently well reproduced
if the phonon spectrum is assumed to be bulk-like.

We will start discussing the scattering-induced damping of 
Bloch oscillations. In particular, we will show that in the low-density limit 
this damping is mainly determined by optical-phonon 
scattering \prc{Rossi95a,Rossi95b} 
while at high densities the main mechanism
responsible for the suppression of Bloch 
oscillations is found to be carrier-carrier scattering \prc{Rossi96}.

This Bloch-oscillation analysis in the time domain is also confirmed by its
counterpart in the frequency domain. As pointed out in 
Sect.~\prr{s:qmp}, 
the presence of Bloch oscillations, due to a negligible scattering 
dynamics, should correspond to Wannier-Stark energy quantization.
This is confirmed by the simulated optical-absorption spectra, 
which clearly show the presence of the field-induced Wannier-Stark ladders 
introduced in Sect.~\prr{ss:wsp} \prc{Koch95}.
\subsection{Bloch-oscillation analysis}\label{ss:boa}

All the simulated experiments presented in this section refer to 
the superlattice structure considered in Ref.~\prc{Meier95b}:
$111$ \AA\ GaAs wells and $17$ \AA\ Al$_{0.3}$Ga$_{0.7}$As barriers. 
For such a structure
there has been experimental evidence for a THz-emission 
from Bloch oscillations \prc{Roskos94}.

In the first set of simulated experiments an initial distribution of 
photoexcited carriers (electron-hole pairs) is generated by a
$100$ fs Gaussian laser pulse in resonance with the first-miniband 
exciton ($\hbar \omega_L \approx 1540$ meV).
The strength of the applied electric field is assumed to be $4$ kV/cm, which 
corresponds to a Bloch period $\tau_B = h/eFd$ of about $800$ fs.

In the low-density limit (corresponding to a weak laser excitation),
incoherent scattering processes do not alter the Bloch-oscillation dynamics. 
This is due to the following reasons:
In agreement with recent experimental \prc{Roskos94,vonPlessen94} and 
theoretical \prc{Rossi95a,Rossi95b,Meier95b} investigations,
for superlattices characterized by a miniband width
smaller than the LO-phonon energy 
---as for the structure considered here---
and for laser 
excitations close to
the band gap, at low temperature carrier-phonon scattering is not 
permitted. 
Moreover, 
in this low-density regime carrier-carrier scattering plays no role: Due 
to the quasi-elastic nature of Coulomb collisions, 
in the low-density limit the majority of the 
scattering processes 
is characterized by a very small
momentum transfer; As a consequence, the momentum relaxation along the 
growth direction is negligible.
As a result, on this picosecond time-scale the carrier system exhibits a 
coherent Bloch-oscillation dynamics, i.e. a negligible scattering-induced 
dephasing.
This can be clearly seen from the time evolution of the carrier distribution 
as a function of $k_\parallel$
(i.e. averaged over $k_{\perp}$) 
shown in Fig.~\prr{fig3}. 
During the laser photoexcitation ($t = 0$) 
the carriers are generated
around $k_\parallel = 0$, where the transitions are close to resonance with
the laser excitation. According to the
acceleration theorem (\prr{eq0}), the electrons are then shifted in
${\bf k}$-space. When the carriers reach the border of the first
Brillouin zone they are Bragg reflected.
After about $800$ fs, corresponding to the Bloch period $\tau_B$, 
the carriers have completed one oscillation
in ${\bf k}$-space. 
As expected, the carriers execute Bloch oscillations without
loosing the synchronism of their motion by scattering.
This is again shown in Fig.~\prr{fig3}, where we have plotted: (b) the mean 
kinetic energy, (c) the current, and (d) its time derivative
which is proportional to the emitted far field, i.e. the THz-radiation.
(It can be shown that, by neglecting Zener tunneling,  the intraband 
polarization $P^{e/h}$ in Eq.~(\prr{eq01}) is proportional to the current.)
All these three quantities exhibit oscillations characterized by the same 
Bloch period $\tau_B$.
Due to the finite width of the carrier distribution in ${\bf k}$-space 
(see Fig.~\prr{fig3}(a)), 
the amplitude of the oscillations of the kinetic energy is somewhat smaller 
than the miniband width. 
Since for this excitation condition the 
scattering-induced dephasing is negligible,
the oscillations of the current are symmetric around zero, which implies 
that the time average of the current is equal to zero, i.e. no dissipation.

As already pointed out, this ideal Bloch-oscillation regime is typical of a
laser excitation close to gap in the low-density limit. Let us now discuss, 
still at low densities, the case of a laser photoexcitation high in the 
band.
Figure \prr{fig4}(a) shows the THz-signal as obtained from a set of simulated 
experiments corresponding to different laser excitations \prc{Meier95b}.
The different traces correspond to the emitted
THz-signal for increasing excitation energies. We clearly notice the 
presence of Bloch oscillations in all cases. However, the oscillation 
amplitude and decay (effective damping) is excitation-dependent.

For the case of a laser excitation resonant with the first-miniband exciton 
considered above (see Fig.~\prr{fig3}),
we have a strong THz-signal. 
The amplitude of the signal decreases when the excitation energy is
increased. Additionally,
there are also some small changes in the phase of the oscillations,
which are induced by the electron-LO phonon scattering.

When the laser energy comes into resonance with the transitions between the
second electron and hole minibands ($\hbar \omega_L \approx 1625$ meV),
the amplitude of the THz-signal increases again. The corresponding 
THz-transients show an initial part, which is strongly damped
and some oscillations for longer times that are much less damped.
For a better understanding of these results,
we show in Fig.~\prr{fig4}(b) the individual
THz-signals, originating from
the two electron and two heavy-hole minibands for the excitation with
$\hbar \omega = 1640$ meV. 
The Bloch oscillations performed by the electrons within the second miniband
are strongly damped due to intra- and interminiband LO-phonon scattering
processes \prc{Rossi95a,Meier95b}.
Since the width of this second miniband ($45$ meV) is somewhat larger than the
LO-phonon energy, also intraminiband scattering is possible, whenever the
electrons are accelerated into the high-energy region
of the miniband.  
The THz-signal originating from electrons within the first miniband
shows an oscillatory behavior with a small amplitude and a phase 
which is determined by the time the electrons need to relax
down to the bottom of the band.

At the same time, 
the holes in both minibands exhibit undamped Bloch oscillations,
since the minibands are so close in 
energy that for these excitation conditions no LO-phonon emission
can occur. The analysis shows that
at early times the THz-signal is mainly determined by
the electrons within the
second miniband. At later times the observed signal is
due to heavy holes and electrons within the first miniband.

The above theoretical analysis closely resembles 
experimental observations obtained for a superlattice structure 
very similar to the one modelled here \prc{Roskos94}. 
In these experiments, evidence for THz-emission 
from Bloch oscillations has been reported. For some 
excitation conditions these oscillations are associated
with resonant excitation of the second miniband. 
The general behavior of the
magnitude of the signals, the oscillations and the damping
are close to the results shown in Fig.~\prr{fig4}.

Finally, in order to study the density dependence of the Bloch-oscillation 
damping, let us go back to the case of laser excitations close to gap.
Figure \prr{fig5}(a) shows the total (electrons plus holes) THz-radiation 
as a function of time for three different carrier densities. 
With increasing carrier density, carrier-carrier scattering
becomes more and more important: Due to Coulomb screening, the 
momentum transfer in a carrier-carrier scattering increases (its typical 
value being comparable with the screening wavevector). This can be seen 
in Fig.~\prr{fig5}(a), where for increasing carrier densities we realize an 
increasing damping of the THz-signal. 
However, also for the highest carrier density considered here we deal with 
a damping time of the order of $700$ fs, which is much larger than the 
typical dephasing time, i.e. the decay time of the interband polarization,
 associated with carrier-carrier scattering.
The dephasing time is typically 
 investigated by means of four-wave-mixing (FWM)
measurements and such multi-pulse experiments can be simulated as 
well \prc{Lohner93,Leitenstorfer94a}.
From a theoretical point of view, a qualitative estimate of the dephasing 
time is given by the decay time of the ``incoherently summed'' polarization 
(ISP) \prc{Kuhn92b}.
Figure \prr{fig5}(b) 
shows such ISP as a function of time for the same three carrier 
densities of Fig.~\prr{fig5}(a).
As expected, the decay times are always much smaller than the 
corresponding damping times of the THz-signals (note the 
different time-scale in Figs.~\prr{fig5}(a) and (b)).
This difference can be understood as follows:
The fast decay times of Fig.~\prr{fig5}(b) reflect the 
interband dephasing, i.e. the
sum of the electron and hole scattering rates. In particular, for the 
Coulomb interaction this means the sum of electron-electron, electron-hole,
and hole-hole scattering.
This last contribution is known to dominate and determines the dephasing 
time-scale.
On the other hand, the total THz-radiation in Fig.~\prr{fig5}(a) 
is the sum of the 
electron and hole contributions. However, due to the small value of the 
hole miniband width compared with the electron one, the electron 
contribution will dominate.
This is clearly shown in Fig.~\prr{fig6}, where the electron (a) and hole (b) 
contributions to the THz-radiation are plotted as a function of time (note 
the different vertical scale). 
This means that the THz damping in Fig.~\prr{fig5}(a) mainly reflects 
the damping of the electron contribution 
(see Fig.~\prr{fig6}(a)).
This decay, in turn, reflects the intraband dephasing of electrons which is
due to electron-electron and electron-hole scattering only, i.e. no hole-hole 
contributions.
This clearly explains the different decay times of Fig.~\prr{fig5}(a) and 
Fig.~\prr{fig5}(b).

From the above analysis we can conclude that the decay time 
of the THz-radiation due to carrier-carrier scattering differs considerably
from the corresponding dephasing times obtained from a FWM experiment:
The first one is a measurement of the intraband dephasing while the second 
one reflects the interband dephasing.
\subsection{Optical-absorption analysis}\label{ss:oaa}

Let us now discuss the frequency-domain counterpart of the 
Bloch-oscilla\-tion picture considered so far.
Similar to what happens in the time domain, for sufficiently high 
electric fields, i.e. when the
Bloch period $\tau_B = h/eFd$ becomes smaller than
the dephasing time, 
the optical spectra of the superlattice are expected to exhibit the 
frequency-domain counterpart of the
Bloch oscillations, i.e. the Wannier-Stark energy quantization discussed in
Sect.~\prr{ss:wsp}.
In absence of Coulomb interaction, the Wannier-Stark ladder absorption 
increases as a function of the photon energy in a step-like fashion.
These steps are equidistantly spaced. This spacing, named Wannier-Stark 
splitting, is proportional to the applied electric field (see 
Eq.~(\prr{eq75})).

The simulated linear-absorption spectra corresponding to a superlattice 
structure with 
$95$ \AA\ GaAs wells and $15$ \AA\ Al$_{0.3}$Ga$_{0.7}$As barriers 
are shown in Fig.~\prr{fig7} \prc{Koch95}.
As we can see, 
the Coulomb interaction gives rise to excitonic peaks in the absorption 
spectra and introduces couplings between these Wannier-Stark states.
Such exciton peaks, which are not equidistantly spaced any more, 
are often referred to as  
excitonic Wannier-Stark ladders \prc{Dignam90} of the 
superlattice. 

Since for the superlattice structure considered in this simulated 
experiment \prc{Rossi95a,Koch95} the combined 
miniband width is larger than the typical 
two- and three-dimensional exciton binding energies,
it is possible to investigate the quasi-three 
dimensional absorption behavior
of the delocalized miniband states as well as localization
effects induced by the electric field.

For the free-field case, the electron and hole states are completely
delocalized in our three-dimensional ${\bf k}$-space.
The perturbation induced by the application of a low field 
(here $\approx 5$ kV/cm), couples the
states along the field direction and in the spectra
the Franz-Keldysh effect,
well known from bulk materials, appears:
one clearly notices oscillations which increase in amplitude with the field
and shift with $F^{2/3}$ from the $n = 0$ and $n = 1$ levels
toward the center of the combined miniband. 

For increasing field the potential drop over the distance of a
few quantum wells eventually exceeds the miniband width and the 
electronic states 
become more and more localized. Despite the field-induced
energy difference $n eFd$,
the superlattice potential is equal for quantum wells separated by $nd$. 
Therefore,
the spectra decouple into a series of peaks
corresponding to the excitonic ground states of the individual
electron-hole Wannier-Stark levels.
Each Wannier-Stark transition contributes to the absorption
with a pronounced $1-s$ exciton peak, plus
higher bound exciton and continuum states. The oscillator
strength of a transition $n$ is proportional
to the overlap between electron and hole wavefunctions centered
at quantum wells $n'$ and $n+n'$, respectively. 
The analysis shows that this oscillator strength
is almost exclusively determined by the amplitude of the electron
wavefunction in quantum well $n'+n$ since for fields in the
Wannier-Stark regime
the hole wavefunctions are almost completely localized over one
quantum well due to their high effective mass
(see Fig.~\prr{fig2}). Thus, the
oscillator strengths of transitions
to higher $|n|$ become smaller with increasing $|n|$ and field.    

At high fields (here $>\approx 8$ kV/cm) the separation between the peaks 
is almost equal to $neFd$.
For example, the peak of the $n=0$ transition
which is shifted by the 
Wannier-Stark exciton binding energy with respect 
to the center of the combined miniband, demonstrates that the
increasing localization also increases the 
exciton binding energy.  This increased 
excitonic binding reflects the gradual transition from
a three- to a two-dimensional behavior. 

For intermediate fields there is an interplay between the
Wannier-Stark and the Franz-Keldysh effect.
Coming from high fields, first the Wannier-Stark peaks are
modulated by the Franz-Keldysh oscillations.
However, as soon as the separation $eFd$ between neighboring peaks becomes 
smaller than their spectral widths,
the peaks can no longer be resolved individually so that only
the Franz-Keldysh structure remains.   
\section*{Acknowledgments}
I wish to thank the colleagues of the Marburg group, S.W. Koch, 
T. Meier, and P. Thomas, for their essential contributions to the 
research activity reviewed in this chapter. 
I am also grateful to A. Di Carlo, M. Gulia, T. Kuhn, E. Molinari, and 
P.E. Selbmann for stimulating and fruitful discussions. 

This work was supported in part by the EC Commission through the Network 
``ULTRAFAST''.
\newpage
\begin{figure}
\section*{Figure captions}
\caption{\label{fig1}
Schematic illustration of the field-induced coherent motion of 
an electronic wavepacket initially created at the bottom of a miniband. 
Here, the width of the miniband exceeds the LO-phonon energy $E_{LO}$, 
so that LO-phonon scattering is possible. After Ref.~\prc{vonPlessen94}.}
\caption{\label{fig2}
Schematic representation of the transitions from the valence to the 
conduction band of a superlattice in the Wannier-Stark localization regime. 
After Ref.~\prc{Waschke93}.}
\caption{\label{fig3}
Full Bloch-oscillation dynamics corresponding to a laser 
photoexcitation resonant with the first-miniband exciton. (a) Time 
evolution of the electron distribution as a function of $k_\parallel$.
(b) Average kinetic energy, (c) current, and (d) THz-signal corresponding 
to the Bloch oscillations in (a).}
\caption{\label{fig4}
(a) Total THz-signals for eight different spectral positions of 
the exciting laser pulse: $1540, 1560,\dots, 1680$ meV (from bottom to top).
(b) Individual THz-signal of the electrons and holes in the different bands
for a central spectral position of the laser pulse of 1640 meV.
After Ref.~\prc{Koch95}.}
\caption{\label{fig5}
(a) Total THz-radiation as a function of time;
(b) Incoherently-summed polarization as a function of time. After 
Ref.~\prc{Rossi96}.} 
\caption{\label{fig6}
(a) Electron and (b) hole contributions to the total THz-radiation 
of Fig.~\prr{fig5}(a). After Ref.~\prc{Rossi96}.}
\caption{\label{fig7}
Absorption spectra for various static applied electric fields for a
GaAs/Al$_{0.3}$Ga$_{0.7}$As superlattice (well (barrier) width $95$ ($15$) 
\AA). The vertical displacements between any two spectra is proportional 
to the difference of the corresponding fields.
The Wannier-Stark transitions are labeled by numbers, the lower (higher) 
edge of the combined miniband by $E_0$ ($E_1$).
After Ref.~\prc{Koch95}.}
\end{figure}
\end{document}